\documentclass[sigconf]{acmart}
\usepackage{xcolor}
\usepackage{colortbl}
\usepackage{amsfonts}
\usepackage{pifont}
\usepackage{multirow}
\usepackage{amsmath}
\usepackage{enumitem}
\usepackage{algorithm}
\usepackage{algorithmic}
\usepackage{upgreek}
\usepackage{mathtools}

\AtBeginDocument{%
  }

\copyrightyear{2025}
\acmYear{2025}
\setcopyright{acmlicensed}\acmConference[WWW '25]{Proceedings of the ACM Web Conference 2025}{April 28-May 2, 2025}{Sydney, NSW, Australia}
\acmBooktitle{Proceedings of the ACM Web Conference 2025 (WWW '25), April 28-May 2, 2025, Sydney, NSW, Australia}
\acmDOI{10.1145/3696410.3714873}
\acmISBN{979-8-4007-1274-6/25/04}

\begin{document}

\title{Hyperbolic Diffusion Recommender Model}
\author{Meng Yuan}
\authornote{Equal contribution.}
\affiliation{%
  \institution{Institute of Artificial Intelligence, Beihang University}
  \city{Beijing}
  \country{China}}
\email{yuanmeng97@buaa.edu.cn}

\author{Yutian Xiao}
\authornotemark[1] 
\affiliation{%
 \institution{Institute of Artificial Intelligence, Beihang University}
 \city{Beijing}
 \country{China}}
\email{by2442221@buaa.edu.cn}

\author{Wei Chen}
\affiliation{%
  \institution{Institute of Artificial Intelligence, Beihang University}
  \city{Beijing}
  \country{China}}
\email{chenwei23@buaa.edu.cn}

\author{Chu Zhao}
\affiliation{%
 \institution{Chongqing University of Technology}
 \city{Chongqing}
 \country{China}}
\email{zhaochu123@2019.cqut.edu.cn}

\author{Deqing Wang}
\affiliation{%
  \institution{School of Computer Science and Engineering, Beihang University}
  \city{Beijing}
  \country{China}}
\email{dqwang@buaa.edu.cn}

\author{Fuzhen Zhuang}
\authornote{Corresponding author.}
\authornote{Fuzhen Zhuang is also at Zhongguancun Laboratory, Beijing, China}
\affiliation{%
  \institution{Institute of Artificial Intelligence, Beihang University}
  \city{Beijing}
  \country{China}}
\email{zhuangfuzhen@buaa.edu.cn}

\renewcommand{\shortauthors}{Yuan et al.}

\begin{abstract}
Diffusion models (DMs) have emerged as the new state-of-the-art family of deep generative models. To gain deeper insights into the limitations of diffusion models in recommender systems, we investigate the fundamental structural disparities between images and items. Consequently, items often exhibit distinct anisotropic and directional structures that are less prevalent in images. However, the traditional forward diffusion process continuously adds isotropic Gaussian noise, causing anisotropic signals to degrade into noise, which impairs the semantically meaningful representations in recommender systems.

Inspired by the advancements in hyperbolic spaces, we propose a novel \textit{\textbf{H}yperbolic} \textit{\textbf{D}iffusion} \textit{\textbf{R}ecommender} \textit{\textbf{M}odel} (named HDRM). Unlike existing directional diffusion methods based on Euclidean space, the intrinsic non-Euclidean structure of hyperbolic space makes it particularly well-adapted for handling anisotropic diffusion processes. In particular, we begin by formulating concepts to characterize latent directed diffusion processes within a geometrically grounded hyperbolic space. Subsequently, we propose a novel hyperbolic latent diffusion process specifically tailored for users and items. Drawing upon the natural geometric attributes of hyperbolic spaces, we impose structural restrictions on the space to enhance hyperbolic diffusion propagation, thereby ensuring the preservation of the intrinsic topology of user-item graphs. Extensive experiments on three benchmark datasets demonstrate the effectiveness of HDRM. Our code is available at
\url{https://github.com/yuanmeng-cpu/HDRM}.

\end{abstract}


\maketitle

\section{Introduction}
Diffusion models (DMs) \cite{ho2020denoising, sohl2015deep, song2019generative, song2020score} have emerged as the new state-of-the-art family of deep generative models. They have broken the long-time dominance of generative adversarial networks (GANs) \cite{goodfellow2014generative} in the challenging task of image synthesis \cite{dhariwal2021diffusion, ho2020denoising, song2020score} and have demonstrated promise in computer vision, ranging from video generation~\cite{ho2022imagen, ho2022video}, semantic segmentation~\cite{baranchuklabel, tan2022semantic}, point cloud completion~\cite{luo2021diffusion, zheng2024point} and anomaly detection~\cite{wyatt2022anoddpm, zhang2023unsupervised}.

\begin{figure*}[t]
	 \centering
	\begin{minipage}{0.235\linewidth}
		\vspace{10pt}
\centerline{\includegraphics[width=1.0\textwidth]{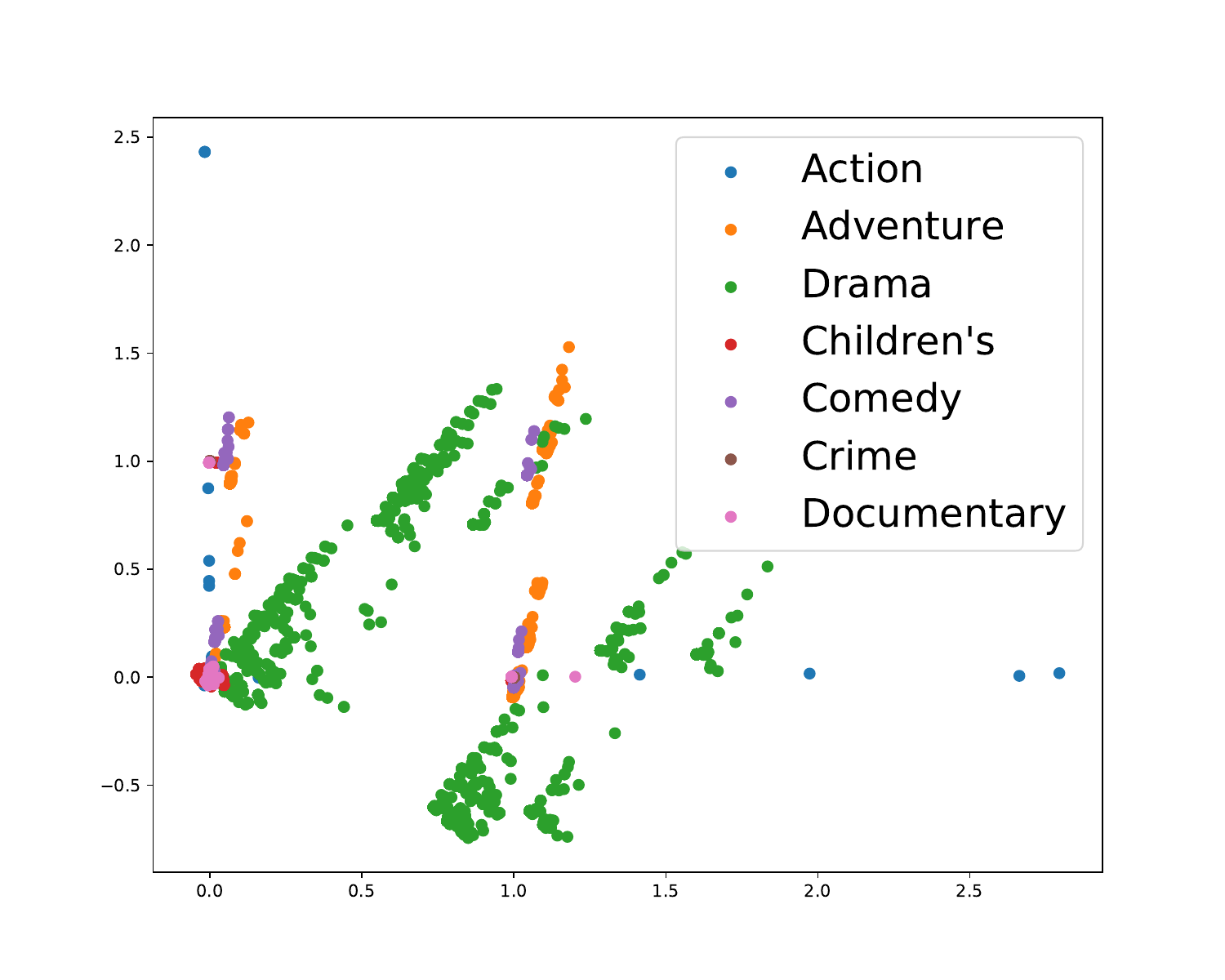}}
		\centerline{(a) ML-1M (Euclidean space)}
	\end{minipage}
 \begin{minipage}{0.25\linewidth}
		\vspace{10pt}
\centerline{\includegraphics[width=1.0\textwidth]{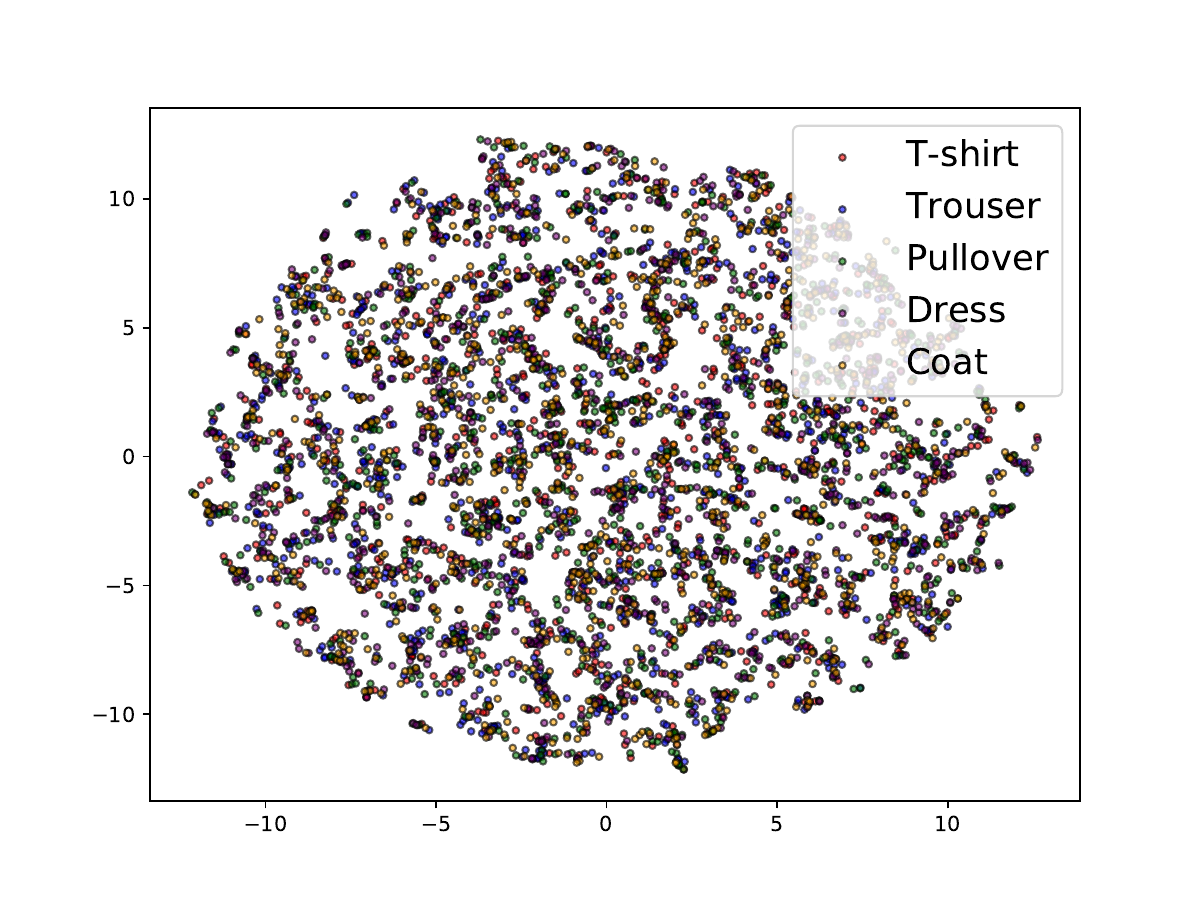}}
		\centerline{(b) F-MNIST (Euclidean space)}
	\end{minipage}
  	\begin{minipage}{0.25\linewidth}
		\vspace{10pt}
\centerline{\includegraphics[width=1.\textwidth]{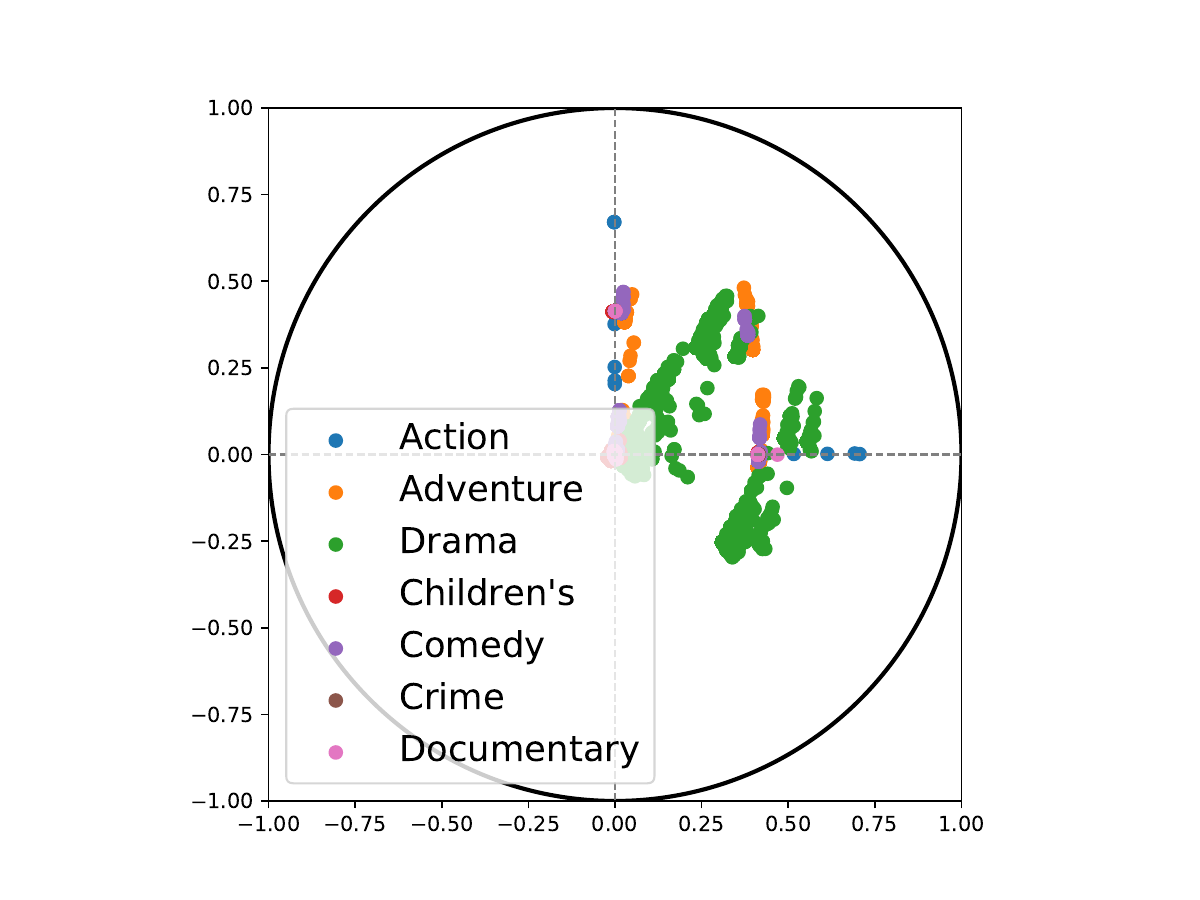}}
		\centerline{(c) ML-1M (Poincaré disk)}
	\end{minipage}
  	\begin{minipage}{0.249\linewidth}
		\vspace{10pt}
\centerline{\includegraphics[width=1.\textwidth]{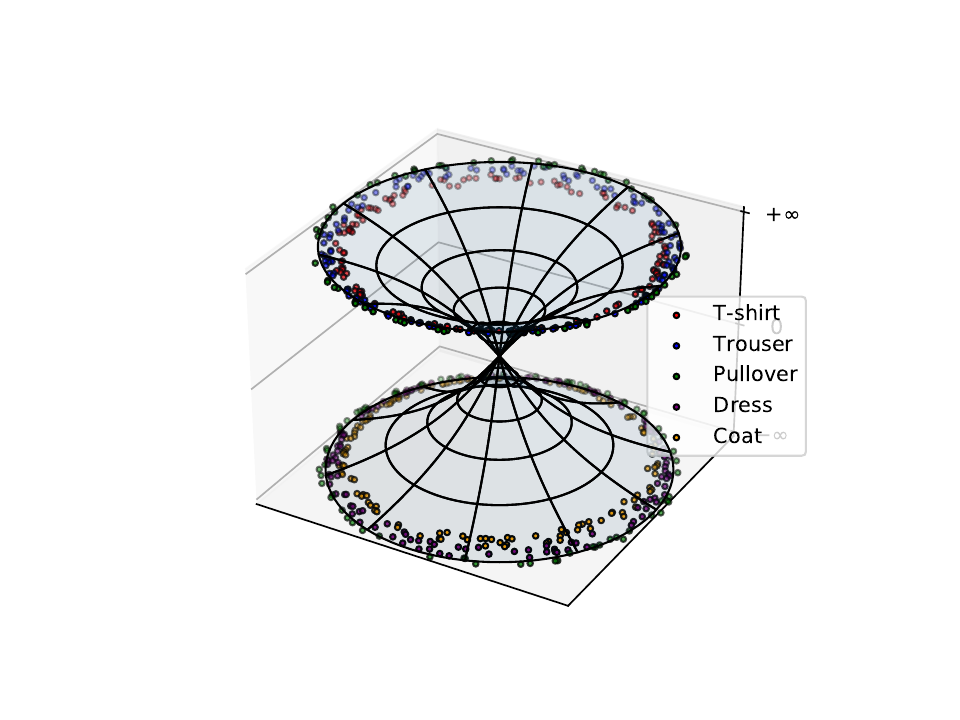}}
		\centerline{(d) F-MNIST (Lorentz manifold)}
	\end{minipage}
	\caption{2D visualization of the data using SVD decomposition, where each color corresponds to a unique category. (a) Euclidean visualization of the item features in MovieLens-1M; (b) Euclidean visualization of the image features in Fashion-MNIST; (c) Hyperbolic visualization of the item features in MovieLens-1M; (d) Hyperbolic visualization of the image features in Fashion-MNIST.}
	\label{fig1}
\end{figure*}

Despite the increasing research on diffusion models in computer vision~\cite{rombach2022high, lu2022conditional, tashiro2021csdi, dhariwal2021diffusion, ho2020denoising, song2020score}, their potential in recommender systems has not been equally explored. Generative recommender models~\cite{liang2018variational, yu2019vaegan, zhao2024graph, zhao2024symmetric, zhao2024denoising, wu2016collaborative} aim to align with the user-item interaction generation processes observed in real-world environments. Unlike other earlier generative recommender models like VAEs~\cite{liang2018variational, wu2016collaborative} and GANs~\cite{wang2017irgan, yu2019vaegan}, 
diffusion recommender models~\cite{wang2023diffusion, zhao2024denoising, li2024multi} leverage a denoising framework to effectively reverse a multi-step noising process to generate synthetic data that matches closely with the distribution of the training data. This highlights the exceptional ability of diffusion models to capture multi-scale feature representations and generate high-quality samples, while also ensuring improved stability during training. However, the aforementioned diffusion recommender models are still directly based on extensions of computer vision methods, neglecting the latent structural differences between images and items.

To gain deeper insights into the limitations of traditional diffusion models in recommender systems, we begin by investigating the fundamental structural disparities between images and items. Specifically, we apply singular value decomposition~\cite{yang2024directional} to both image and graph data, and plot the resulting projections on a two-dimensional plane. Figure\textcolor{red}{~\ref{fig1}a} reveals that the projected data from ML-1M exhibits strong anisotropic structures across multiple directions, whereas the projected images from F-MNIST (as seen in Figure\textcolor{red}{~\ref{fig1}b}) form a relatively more isotropic distribution centered around the origin. As a result, items often exhibit distinct anisotropic and directional structures that are less prevalent in images~\cite{yi2024directional}. Unfortunately, the traditional forward diffusion process continuously adds isotropic Gaussian noise, causing anisotropic signals to degrade into noise~\cite{yang2024directional}, which impairs the semantically meaningful representations in recommender systems.

Hyperbolic spaces are extensively regarded as the optimal continuous manifold for modeling discrete tree-like or hierarchical structures~\cite{bachmann2020constant, sala2018representation, sun2024motif, krioukov2010hyperbolic}, and have been widely studied and applied to various recommender tasks~\cite{chen2022modeling, tai2021knowledge, yuan2023knowledge, yuan2024light, vinh2020hyperml, sun2021hgcf, yang2022hicf}. In hyperbolic spaces, the expansion of space is not uniform (i.e., isotropic), but rather depends on the position and direction. This leads to variations in the rate of change in distances between points along different directions. As shown in Figure\textcolor{red}{~\ref{fig1}c}, hyperbolic spaces are well-suited to preserving the anisotropy of data due to its inherent geometric properties. Additionally, due to the infinite volume of hyperbolic space~\cite{sala2018representation,nickel2017poincare}, modeling uniformly distributed data tends to push the data features toward the boundary, thereby weakening the isotropy of the data to some extent (as seen in Figure\textcolor{red}{~\ref{fig1}d}).


Inspired by the advancements in hyperbolic spaces, we propose a novel \textit{\textbf{H}yperbolic} \textit{\textbf{D}iffusion} \textit{\textbf{R}ecommender} \textit{\textbf{M}odel} named HDRM. Unlike existing directional diffusion methods based on Euclidean space~\cite{yang2024directional, yi2024directional}, the intrinsic non-Euclidean structure of hyperbolic space makes it particularly well-adapted for handling anisotropic diffusion processes. In particular, we begin by formulating concepts to characterize latent directed diffusion processes within a geometrically grounded hyperbolic space. Subsequently, we propose a novel hyperbolic latent diffusion process specifically tailored for users and items. Drawing upon the natural geometric attributes of hyperbolic spaces, we impose structural restrictions on the space to execute hyperbolic preference directional diffusion, thereby ensuring the preservation of the intrinsic topology of user-item graphs. Extensive experiments on three benchmark datasets demonstrate the effectiveness of HDRM. To summarize, we highlight the key contributions of this paper as follows:
\begin{itemize}[leftmargin=*]
\item We contribute to the exploration of anisotropic structures in recommender systems. To the best of our knowledge, this is the first work to design a hyperbolic diffusion model for recommender systems.

\item We propose a novel hyperbolic latent diffusion
process specifically tailored for users and items. Drawing upon the natural geometric attributes of hyperbolic spaces, we impose structural restrictions to facilitate directional diffusion propagation.


\item Extensive experimental results on three benchmark datasets demonstrate that HDRM outperforms various baselines. Further ablation studies verify the importance of each module.
\end{itemize}

\section{PRELIMINARIES}
This section provides foundational concepts, including hyperbolic spaces and diffusion models, to aid in the reader's understanding.

\subsection{Hyperbolic Spaces}
Here we introduce some fundamental concepts of hyperbolic spaces. For more detailed operations on hyperbolic spaces, please refer to Appendix \ref{Hyperbolic Spaces}.

\begin{itemize}[leftmargin=*]
\item
\textbf{Manifold}: Consider a manifold $\mathcal{M}$ with $n$ dimensions as a space where the local neighborhood of a point can be closely approximated by Euclidean spaces $\mathbb{R}^{n}$. For instance, the Earth can be represented by a spherical space, its immediate vicinity can be approximated by $\mathbb{R}^{2}$.

\item\textbf{Tangent space}: For every point $x\in\mathcal{M}$, the tangent space $\mathcal{T}_{x}\mathcal{M}$ of $\mathcal{M}$ at $x$ is set as a $n$-dimensional space measuring $\mathcal{M}$ around x at a first order. 

\item\textbf{Geodesics distance}: This denotes the generalization of a straight line to curved spaces, representing the shortest distance between two points within the context of the manifold.

\item\textbf{Exponential map}: The exponential map carries a vector $v\in\mathcal{T}_{x}\mathcal{M}$ of a point $x\in\mathcal{M}$ to the manifold $\mathcal{M}$, i.e., $\exp_{x}^{\kappa}: \mathcal{T}_{x}\mathcal{M}\rightarrow \mathcal{M}$ by simulating a fixed distance along the geodesic defined as $\gamma(0)=x$ with direction $\gamma'(0)=v$. Each manifold corresponds to its unique way of constructing exponential maps.

\item\textbf{Logarithmic map}: Serving as the counterpart to the exponential map, the logarithmic map takes a point $z$ from the manifold $\mathcal{M}$ and maps it back to the tangent space $\mathcal{T}_{x}\mathcal{M}$, i.e., $\log_{x}^{\kappa}: \mathcal{M} \rightarrow \mathcal{T}_{x}\mathcal{M}$. Like $\exp_{x}^{\kappa}$, each manifold has its formula that defines $\log_{x}^{\kappa}$.
\end{itemize}

\subsection{Diffusion Models}
DMs have attained remarkable success across numerous domains, primarily through the use of forward and reverse processes~\cite{ wang2023diffusion, rombach2022high}.
\begin{itemize}[leftmargin=*]
    \item \textbf{Forward Process}: Given an input data sample \( x_0 \sim q(x_0) \), the forward process constructs the latent variables \( x_{1:T} \) by gradually adding Gaussian noise in \( T \) steps. Specifically, DMs define the forward transition \( x_{t-1} \to x_t \) as:
\begin{equation}
\begin{aligned}
q(x_t|x_{t-1}) &= \mathcal{N}(x_t; \sqrt{1 - \beta_t}x_{t-1}, \beta_t \mathrm{I}),\\
&= \sqrt{1 - \beta_t}x_{t-1} + \sqrt{\beta_t}\epsilon, \quad \epsilon \sim \mathcal{N}(0, \mathrm{I})
\end{aligned}
\label{eq1}
\end{equation}
where \( t \in \{1, \dots, T\} \) represents the diffusion step, \( \mathcal{N}(0, \mathrm{I}) \) denotes the Gaussian distribution, and \( \beta_t \in (0, 1) \) controls the amount of noise added at each step. This method shows the flexibility of the direct sampling of $x_t$ conditioned on the input $x_{t-1}$ at an arbitrary diffusion step $t$ from a random Gaussian noise $\epsilon$.
\item \textbf{Reverse Process}: DMs learn to remove the noise from \( x_t \) to recover \( x_{t-1} \) in the reverse process, aiming to capture subtle changes in the generative process. Formally, taking \( x_T \) as the initial state, DMs learn the denoising process \( x_t \to x_{t-1} \) iteratively as follows:
\begin{equation}
\begin{aligned}
p_\theta(x_{t-1}|x_t) = \mathcal{N}(x_{t-1}; \mu_\theta(x_t, t), \Sigma_\theta(x_t, t)),
\end{aligned}
\label{eq2}
\end{equation}

where \( \mu_\theta(x_t, t) \) and \( \Sigma_\theta(x_t, t) \) are the mean and covariance of the Gaussian distribution predicted by parameters \( \theta \).

\item \textbf{Optimization}: For training the diffusion models, the key focus is obtaining reliable values for $\mu_\theta(x_t, t)$ and $\Sigma_\theta(x_t, t)$ to guide the reverse process towards accurate denoising. To achieve this, it is important to optimize the variational lower bound of the negative log-likelihood of the model’s predictive denoising distribution $p_\theta(x_0)$:
\begin{equation}
\begin{aligned}
\mathcal{L} &= \mathbb{E}_{q(x_0)} [-\log p_\theta(x_0)] \\
&\leq \mathbb{E}_q [L_\mathrm{T} + L_{\mathrm{T}-1} + \dots + L_0],
\quad \text{where}\\
\label{eq3}
\end{aligned}
\end{equation}
\vspace{-5
mm}
\begin{equation}
\begin{split}
L_\mathrm{T} &= D_{\text{KL}}(q(x_\mathrm{T} | x_0) \parallel p_\theta(x_\mathrm{T})),\\
L_t &= D_{\text{KL}}(q(x_t | x_{t+1}, x_0) \parallel p_\theta(x_t | x_{t+1})),\\
L_0 &= -\log p_\theta(x_0 | x_1),\\
\end{split}
\label{eq4}
\end{equation}
where $t \in \{1, 2, \dots, \mathrm{T} - 1\}$. While $L_\mathrm{T}$ can be disregarded during training due to the absence of learnable parameters in the forward process, $L_0$ represents the negative log probability of the original data sample $x_0$ given the first-step noisy data $x_1$, and $L_t$ aims to align the distribution $p_\theta(x_t | x_{t+1})$ with the tractable posterior distribution $q(x_t | x_{t+1}, x_0)$ in the reverse process \cite{luo2022understanding}.

\item \textbf{Inference}: After training the model parameters \( \theta \), DMs can sample \( x_T \) from a standard Gaussian distribution \( \mathcal{N}(0, \mathrm{I}) \), and subsequently utilize \( p_\theta(x_{t-1} | x_t) \) to iteratively reconstruct the data, following the reverse process
$x_\mathrm{T} \to x_{\mathrm{T}-1} \to \cdots \to x_0$.
In addition, previous works~\cite{rombach2022high, li2022diffusion} have explored the incorporation of specific conditions to enable controlled generation.
\end{itemize}

\section{METHOD}

\begin{figure}[t]
	\centering
	\setlength{\fboxrule}{0.pt}
	\setlength{\fboxsep}{0.pt}
	\fbox{      \includegraphics[width=1.0\linewidth]{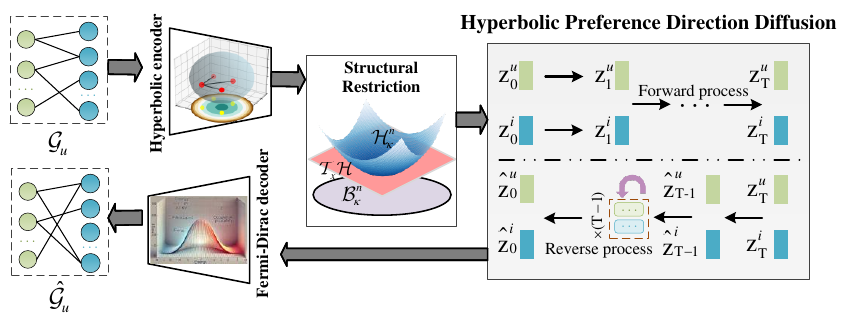}}
	\caption{An overview illustration of the HDRM architecture.}
	\label{fig2}
\end{figure}

In light of the successful applications of diffusion models~\cite{wang2023diffusion, zhao2024denoising, li2024multi}, we employ a two-stage training strategy for our implementation. First, we train the hyperbolic encoder to generate pre-trained user and item embeddings. Subsequently, we proceed with the training of the hyperbolic latent diffusion process. The overall architecture is illustrated in Figure \ref{fig2}.

\subsection{Hyperbolic Geometric Autoencoding}

\subsubsection{Hyperbolic Graph Convolutional Network}
We adopt the hyperbolic graph convolutional network~\cite{chami2019hyperbolic, sun2021hgcf} as the hyperbolic encoder to embed the user-item interaction graph $\mathcal{G}_{u} = (\mathcal{U}, \mathcal{I})$ into a low-dimensional hyperbolic geometric space, thereby enhancing the subsequent graph latent diffusion process. The objective of the hyperbolic encoder is to generate hyperbolic embeddings for users and items. Formally, we use \(\mathbf{x} \in \mathbb{R}^n\) to represent the Euclidean state of users and items. Then the initial hyperbolic state \(\mathbf{e}_{i}^{(0)}\) and \(\mathbf{e}_{u}^{(0)}\) can be obtained by:
\begin{equation}
\begin{aligned}
\mathbf{e}_i^{(0)} = \exp_{\mathbf{o}}^{\kappa}(\mathbf{z}_i^{(0)}),\quad 
\mathbf{e}_u^{(0)} = \exp_{\mathbf{o}}^{\kappa}(\mathbf{z}_u^{(0)}),
\label{project}
\end{aligned}
\end{equation}
\begin{equation}
\begin{aligned}
    \mathbf{z}_i^{(0)} = (0, \mathbf{x}_i), \quad
    \mathbf{z}_u^{(0)} = (0, \mathbf{x}_u),
\end{aligned}
\end{equation}
where \(\mathbf{x}\) is taken from multivariate Gaussian distribution. \(\mathbf{z}^{(0)} = (0, \mathbf{x})\) denotes the operation of inserting the value 0 into the zeroth coordinate of \(\mathbf{x}\) so that \(\mathbf{z}^{(0)}\) can always live in the tangent space of origin. 

Next, the hyperbolic neighbor aggregation is computed by aggregating the representations of neighboring users and items. Given the neighbors \(\mathcal{N}_i\) and \(\mathcal{N}_u\) of \(i\) and \(u\), respectively, the embedding of user \(u\) and \(i\) is updated using the tangent state $\mathbf{z}$ and the \(k\)-th ($k>0$) aggregation is given by:
\begin{equation}
\begin{aligned}
    \mathbf{z}_i^{(k)} &= \mathbf{z}_i^{(k-1)} + \sum_{u \in \mathcal{N}_i} \frac{1}{|\mathcal{N}_i|} \mathbf{z}_u^{(k-1)}, \\
    \mathbf{z}_u^{(k)} &= \mathbf{z}_u^{(k-1)} + \sum_{i \in \mathcal{N}_u} \frac{1}{|\mathcal{N}_u|} \mathbf{z}_i^{(k-1)},
\end{aligned}
\label{eq3}
\end{equation}
where \(|\mathcal{N}_u|\) and \(|\mathcal{N}_i|\) are the number of one-hop neighbors of \(u\) and \(i\), respectively. For high-order aggregation, sum-pooling is applied in these \(k\) tangential states:
\begin{equation}
\begin{aligned}
    \mathbf{z}_i &= \sum_{k} \mathbf{z}_i^{(k)}, \quad
    \mathbf{z}_u = \sum_{k} \mathbf{z}_u^{(k)}.\\
    \mathbf{e}_i &= \exp_{\mathbf{o}}^{\kappa}(\mathbf{z}_i),\quad 
    \mathbf{e}_u = \exp_{\mathbf{o}}^{\kappa}(\mathbf{z}_u).
\end{aligned}
\label{eq4}
\end{equation}

Note that \(\mathbf{z}\) is on the tangent space of origin. For the hyperbolic state, it is projected back to the hyperbolic spaces with the exponential map.

\subsubsection{Hyperbolic Decoder} 

In accordance with these hyperbolic learning models~\cite{chami2019hyperbolic, nickel2017poincare, ganea2018hyperbolic, krioukov2010hyperbolic}, we use the Fermi-Dirac decoder~\cite{sun2021hgcf, yang2022hicf}, a generalization of sigmoid, to estimate the probability of the user clicking on the item:
\begin{equation}
\mathbf{s}(u,i) = \frac{1}{\exp{(d_{\mathcal{L}}^{\kappa}(\hat{\mathbf{e}}_0^{u}, \hat{\mathbf{e}}_0^{i})^2 - q)/t} + 1},
\label{decoder}
\end{equation}
where $d_{\mathcal{L}}^{\kappa}(\cdot, \cdot)$ is the hyperbolic distance as mentioned in Table ~\ref{tab0}, $\kappa$ denotes the curvature, $\hat{\mathbf{e}}_0^{u}$ and $\hat{\mathbf{e}}_0^{i}$ denote the exponential maps of $\hat{\mathbf{z}}_0^{u}$ and $\hat{\mathbf{z}}_0^{i}$ resulting from the reverse process. $q$ and $t$ are hyper-parameters. Here, we slightly abuse the notation for $\exp$: the unindexed $\exp$ refers to the exponential operation, and $\exp_\mathbf{o}^{\kappa}$ denotes the mapping of embeddings from the tangent space to hyperbolic space.

In summary, the workflow of hyperbolic geometric autoencoding is that the output from the encoder's final layer is projected into hyperbolic space through exponential mapping, after which the sampled latent vector is returned to Euclidean space via logarithmic mapping before being passed into the decoder layers.


\subsection{Hyperbolic Preference Directional Diffusion} 
 
Unlike the uniform, isotropic expansion in Euclidean space, the volume of hyperbolic space increases exponentially with radius, reflecting its intrinsic anisotropy. Consequently, a key challenge lies in effectively leveraging this anisotropic structure to achieve controllable and direction-aware diffusion processes~\cite{yang2024directional,yi2024directional}. 

To this end, inspired by recent advances in hyperbolic diffusion models~\cite{lin2023hyperbolic, fu2024hyperbolic}, we aim to achieve precise propagation of user preferences by enforcing structural restrictions that enable directed diffusion in hyperbolic space.

\subsubsection{Hyperbolic Clustering}

To conserve computational resources and memory usage, we follow previous works~\cite{wang2023diffusion, zhao2024denoising, fu2024hyperbolic} by clustering items during the pre-processing stage. Formally, if the entitites $\mathbf{e}_i$ represent the $k$-th cluster, the clustering center representation $\mathbf{z}_i$ in the tangent space of $\mu_k$ can be obtained as follows:
\begin{equation}
    \mathbf{z}_{\mu_{k}}^{i} = \log_{\mu_{k}}^{\kappa}(\mathbf{e}_{i}), \quad \mu_k = \arg \min_{\mu_k} \sum_i d_{\mathcal{L}}^{\kappa}(\mathbf{e}_{i}, \mu_k)^2,
\label{hyper-kmeans}
\end{equation}

where $\mu_k$ denotes the center of the $k$-th cluster, determined through hyperbolic version of k-means. Further exploration of hyperbolic clustering can be found in Appendix~\ref{proof_hyperbolic_clustering}.

\subsubsection{Forward Process of Structural Restrictions} 

Hyperbolic spaces provide a natural and geometric framework for modeling the connection patterns of entities during the process of graph growth~\cite{chami2019hyperbolic}. Our goal is to develop a diffusion model that incorporates hyperbolic stride growth, aligning this growth with the inherent properties of hyperbolic spaces.

To ensure the maintenance of this hyperbolic growth behavior in the tangent space, we employ the following formulas:
\begin{equation}
\begin{aligned}
q(\mathbf{z}_t^{u}|\mathbf{z}_{t-1}^{u}) &= \sqrt{1 - \beta_t}\mathbf{z}_{t-1}^{u} + \sqrt{\beta_t} \epsilon_\mathcal{B} + \delta \tanh ( \sqrt{\kappa} \zeta_{\mathbf{z_{t-1}^{u}}}^{\kappa} / \mathrm{r} ) \mathbf{z}_{t-1}^{u},\\
q(\mathbf{z}_t^{i}|\mathbf{z}_{t-1}^{i}) &= \sqrt{1 - \beta_t}\mathbf{z}_{t-1}^{i} + \sqrt{\beta_t} \epsilon_\mathcal{B} + \delta \tanh ( \sqrt{\kappa} \zeta_{\mathbf{z_{t-1}^{i}}}^{\kappa} / \mathrm{r} ) \mathbf{z}_{t-1}^{i},
\end{aligned}
\label{eq_diffusion}
\end{equation}
where $\delta$ is the stride length that determines the diffusion strength in hyperbolic space, $\mathrm{r}$ is a hyper-parameter to control the speed of stride growth rate, $\epsilon_\mathcal{B}$ follows the Poincaré normal distribution~\cite{mirvakhabova2020performance, nagano2019wrapped}, and $\zeta_{\mathbf{z_{t-1}}}^{\kappa}$ is defined as $1/(\kappa |\mathbf{z_{t-1}}|)$.

Inspired by recent directional diffusion models~\cite{yi2024directional, yang2024directional}, we establish the geodesic direction from the center of each cluster to the anchor \( \mathbf{o} \) as the desired diffusion direction: 
\begin{equation}
\label{equation_sgn}
\begin{aligned}
    \mathbf{a}_{u} = \text{sign}(\log_{\mathbf{o}}^{\kappa} (\mathbf{e}_{\mu_{u}}))* \epsilon_\mathcal{B},\quad
    \mathbf{a}_{i} = \text{sign}(\log_{\mathbf{o}}^{\kappa} (\mathbf{e}_{\mu_{i}}))* \epsilon_\mathcal{B},
\end{aligned}
\end{equation}
where \( \text{sign}(\cdot) \) is used to extract the sign of a real number, returning \( 1 \) for positive values, \( -1 \) for negative values, and \( 0 \) when the value is zero. $\mathbf{a}_{u}$ and $\mathbf{a}_{i}$ represent the angle constrained noise, $\mu$ is the clustering center corresponding to each user $u$ and item $i$. By integrating the above structural restrictions, the geometric diffusion process (\textit{cf.} Eq.~\eqref{eq_diffusion}) can be reformulated as follows:
\begin{equation}
\begin{aligned}
q(\mathbf{z}_t^{u}|\mathbf{z}_{t-1}^{u}) &= \sqrt{1 - \beta_t}\mathbf{z}_{t-1}^{u} + \sqrt{\beta_t} \mathbf{a}_{u} + \delta \tanh ( \sqrt{\kappa} \zeta_{\mathbf{z_{t-1}^{u}}}^{\kappa} / \mathrm{r} ) \mathbf{z}_{t-1}^{u},\\
q(\mathbf{z}_t^{i}|\mathbf{z}_{t-1}^{i}) &= \sqrt{1 - \beta_t}\mathbf{z}_{t-1}^{i} + \sqrt{\beta_t} \mathbf{a}_{i} + \delta \tanh ( \sqrt{\kappa} \zeta_{\mathbf{z_{t-1}^{i}}}^{\kappa} / \mathrm{r} ) \mathbf{z}_{t-1}^{i}.
\end{aligned}
\label{eq_11}
\end{equation}

Consider $\mathbf{z}_t$ denotes the user or item at the $t$-step in the forward diffusion process Eq.~\eqref{eq_11}. Since the forward process adds a fixed amount of noise with a normal distribution at each step, similar to Euclidean space, as \( t \) tends to infinity, the \( \mathbf{z}_t \) will approximate a Poincaré normal distribution:
\begin{equation}
\begin{split}
\mathbf{z}_t = \eta \cdot\mathbf{z}_{t-1} + \epsilon_\mathcal{B}, \; \epsilon_\mathcal{B} \sim \mathcal{N}_\mathcal{B}(0, \mathrm{I}) \\
\implies
    \lim_{t \to \infty} \mathbf{z}_t \sim \mathcal{N}_\mathcal{B} (\delta \mathbf{z}_{t-1}, \mathrm{I}).
\end{split}
\end{equation}
For a more detailed discussion on the Poincaré normal distribution, please refer to the Appendix. \ref{proof_poincare_normal_distribution}.

\subsubsection{Reverse Process} 
After getting noisy user embeddings $\mathrm{z}_T^u$ and noisy item embeddings $\mathrm{z}_T^i$ in the forward process, we follow the standard denoising process~\cite{wang2023diffusion, xu2023geometric} (\textit{cf.} Eq.~\eqref{eq2}) and train a denoising network to simulate the process of reverse diffusion. 
\begin{equation}
\begin{aligned}
p_\theta(\hat{\mathbf{z}}_{t-1}^u|\hat{\mathbf{z}}_t^u) &= \mathcal{N}_\mathcal{B}(\hat{\mathbf{z}}_{t-1}^u; \mu_\theta(\hat{\mathbf{z}}_t^u, t), \Sigma_\theta(\hat{\mathbf{z}}_t^u, t)),\\
p_\psi(\hat{\mathbf{z}}_{t-1}^i|\hat{\mathbf{z}}_t^i) &= \mathcal{N}_\mathcal{B}(\hat{\mathbf{z}}_{t-1}^i; \mu_\psi(\hat{\mathbf{z}}_t^i, t), \Sigma_\psi(\hat{\mathbf{z}}_t^i, t)),
\end{aligned}
\label{eq15}
\end{equation}
where $\hat{\mathbf{z}}_t^u$ and $\hat{\mathbf{z}}_t^i$ are the denoised embeddings in the reverse step $t$, $\theta$ and $\psi$ are the learnable parameters of the user denoising module and the item denoising module correspondingly. These denoising modules are applied iteratively in the reverse process until the generation of the final clean embeddings for the user and item, namely $\hat{\mathbf{z}}_0^u$ and $\hat{\mathbf{z}}_0^i$.

\subsection{Optimization} 
\subsubsection{Hyperbolic Margin-based Ranking Loss}
The margin-based ranking loss has shown to be quite beneficial for hyperbolic recommender methods \cite{sun2021hgcf, yang2022hicf, yuan2023knowledge}. This loss aims to distinguish user-item pairs up to a specified margin into positive and negative samples, once the margin is satisfied the pairs are regarded as well separated. Specifically, for each user $u$ we sample a positive item $i$ and a negative item $j$, and the margin loss is described as:
\begin{equation}
\begin{split}
{\mathcal{L}_\mathbf{Rec}(u,i,j) = max(\underbrace{\mathbf{s}(u,j)}_{push}-\underbrace{\mathbf{s}(u,i)}_{pull}+m, 0)},
\end{split}
\label{eq:Eq_19}
\end{equation}
where the $\mathbf{s}(\cdot)$ denotes the Fermi-Dirac decoder (\textit{cf.} Eq.~\eqref{decoder}), $m$ is the margin between $(u,i)$ and $(u,j)$.  As a result, positive items are pulled closer to user while negative items are pushed outside the margin.

\subsubsection{Reconstruction Loss}
To improve the embedding denoising process, it is crucial to minimize the variational lower bound of the predicted user and item embeddings. Based on the KL divergence derived from the multivariate Gaussian distribution (\textit{cf.} Eq.~\eqref{eq4}), the reconstruction loss of denoising process is stated as follows:
\begin{equation}
\mathcal{L}_{\text{re}}(u, i) = \mathbb{E}_q \left[ -\log p_\theta (\hat{\mathbf{z}}_0^u) - \log p_\psi (\hat{\mathbf{z}}_0^i) \right],
\label{eq17}
\end{equation}
where $\hat{\mathbf{z}}_0^u$ and $\hat{\mathbf{z}}_0^i$ are derived from the final step of Eq. \eqref{eq15}. 

To reduce computational complexity, we follow previous works~\cite{zhao2024denoising} by uniformly sampling t from \{1, 2, ..., T\}
and simplify Eq. \eqref{eq17} into the following equation:
\begin{equation}
    \mathcal{L}_\text{re}(u, i) = ( \mathcal{L}_\text{re}^{u} + \mathcal{L}_\text{re}^{i})/2, \quad \text{where}
\end{equation}
\vspace{-4mm}
\begin{equation}
\begin{aligned}
    \mathcal{L}_\text{re}^{u} &= \mathbb{E}_{t \sim \mathcal{U}(1, \mathrm{T})} \mathbb{E}_{q} \left[ || \mathrm{z}^{u}_0 - \hat{\mathbf{z}}_0^u ||^{2}_{2} \right], \\
    \mathcal{L}_\text{re}^{i} &= \mathbb{E}_{t \sim \mathcal{U}(1, \mathrm{T})} \mathbb{E}_{q} \left[ || \mathrm{z}^{i}_0 - \hat{\mathbf{z}}_0^i ||^{2}_{2} \right].
\end{aligned}
\end{equation}

\subsubsection{Total Loss}

The total loss function of HDRM comprises two parts: a hyperbolic margin-based ranking loss for recommendation, and a reconstruction loss for the denoising process. In summary, the total loss function of HDRM is formulated as follows:
\begin{equation}
\mathcal{L}(u, i, j) = \alpha \cdot\mathcal{L}_\mathbf{Rec}(u, i, j) + (1 - \alpha)\cdot\mathcal{L}_\text{re}(u, i),
\end{equation}
where $\alpha$ is a balance factor to adjust the weight of these two losses. 

To further refine HDRM, we introduce a reweighted loss aimed at improving data cleaning. Following previous works~\cite{wang2021denoising}, we dynamically assign lower weights to instances with lower positive scores:
\begin{equation}
\mathrm{w}(u, i, j) = \text{sigmoid}(\mathbf{s}(u, i))^{\gamma},
\end{equation}
\begin{equation}
\mathcal{L}_\text{total}(u, i, j) = \mathrm{w}(u, i, j)\mathcal{L}(u, i, j),
\label{eq22}
\end{equation}
where $\gamma$ is the reweighted factor which regulates the range of weights, $\mathbf{s}(u, i)$ is obtained from Eq. \eqref{decoder}. Consequently, we redefine the total loss function of HDRM as presented in Eq. \eqref{eq22}.


\subsection{Complexity Analysis}
\label{complexity_analysis}
\subsubsection{Time Complexity}
The time complexity of our model is primarily composed of two phases: 1) Hyperbolic embedding and clustering; 2) Diffusion forward process.
\begin{itemize}[leftmargin=*]

\item \textbf{Hyperbolic embedding and clustering}: We encode each user and item into hyperbolic space using hyperbolic GCN. This process results in $n*d$-dimensional vectors, where n is the total number of users and items. The time complexity of this step is $O(nd)*1(t)$, where $1(t)$ represents the time cost of passing through the neural network. The clustering process has an approximate time complexity of $O(cnd)$, where $c$ denotes the number of cluster categories. 

\item \textbf{Diffusion forward process}: For the forward process of diffusion, a single noise addition step suffices. This step has a time complexity of $O(nd)$. The training of denoising networks incurs a complexity of $O(nd)*1(t)$. 
\end{itemize}

In summary, the overall time complexity for each epoch is $O(1(t) * 2nd) + O((c + 1)nd)$.

\subsubsection{Space Complexity}
In HDRM, we encode users and items in hyperbolic space, representing each as an $n*d$-dimensional vectors. This encoding scheme results in a diffusion scale of $O(hnd)$, where $h$ denotes the total number of user-item interactions.

\section{EXPERIMENTS}
In this section, we conduct a series of experiments to validate HDRM and answer the following key research questions:
\begin{itemize}[leftmargin=*]
\item\textbf{RQ1}: How does HDRM perform compared to baseline models on real-world datasets?
\item\textbf{RQ2}: How does each proposed module contribute to the performance?   


 
\item\textbf{RQ3}: How does HDRM perform in mitigating the effects of noisy data?
\item\textbf{RQ4}: How do hyper-parameters influence the performance of HDRM?
\end{itemize}
\subsection{Experimental Settings}
\subsubsection{Datasets and Evaluation Metrics} 
We evaluate HDRM on three real-world datasets: Amazon-Book\footnote{\url{https://jmcauley.ucsd.edu/data/amazon/}}, Yelp2020\footnote{\url{https://www.yelp.com/dataset/}}, and ML-1M\footnote{\url{https://grouplens.org/datasets/movielens/1m/}}. The detailed statistical information is presented in the Table \ref{tab01}. Across all datasets, interactions rating below 4 classify as false-positive engagements. We follow the data partition rubrics in recent collaborative filtering methods~\cite{rendle2009bpr, he2020lightgcn} and split into three parts (training sets, validation sets, and test sets) with a ratio 7:1:2. Our evaluation of top-$\mathrm{K}$ recommendation efficiency involves the full-ranking protocol, incorporating two popular metrics Recall@$\mathrm{K}$ (R@$\mathrm{K}$) and NDCG@$\mathrm{K}$ (N@$\mathrm{K}$) for which we use $\mathrm{K}$ values of 10 and 20.
\begin{table}[h]
\centering
\caption{Statistics of three datasets under two different settings, where ``C'' and ``N'' represent clean training and natural noise training, respectively. ``Int.'' denotes interactions.}
\setlength{\tabcolsep}{0.55mm}{ 
\begin{tabular}{lcccccc}
\toprule
\textbf{Dataset} & \textbf{\#User} & \textbf{\#Item (C)} & \textbf{\#Int. (C)} & \textbf{\#Item (N)} & \textbf{\#Int. (N)} \\
\midrule
Amazon-Book & 108,822 & 94,949 & 3,146,256 & 178,181 & 3,145,223 \\
Yelp2020 & 54,574 & 34,395 & 1,402,736 & 77,405 & 1,471,675 \\
ML-1M & 5,949 & 2,810 & 571,531 & 3,494 & 618,297 \\
\bottomrule
\end{tabular}
}
\label{tab01}
\end{table}

\begin{table*}[h]
\centering
\small 
\caption{The overall performance evaluation results for the proposed method and compared baseline models on three experimented datasets, highlighting the best and second-best performances in bold and borderline, respectively. Numbers with an asterisk (*) indicate statistically significant improvements over the best baseline (t-test with p-value \textless 0.05).}
 \renewcommand{\arraystretch}{1.1} 

 \setlength{\tabcolsep}{1.mm}{ 
    \begin{tabular}{lcccccccccccccl}
     \toprule
     & \multirow{2}{*}{Model}& \multicolumn{4}{c}{ML-1M}&\multicolumn{4}{c}{Amazon-Book} &\multicolumn{4}{c}{Yelp2020}
     \\
    \cmidrule(r){3-6} \cmidrule(r){7-10} \cmidrule(r){11-14}
    & & R@10 & N@10 &R@20 &N@20 & R@10 & N@10 &R@20 & N@20 & R@10 & N@10 &R@20  &N@20 \\
    \hline
    \hline
    &BPRMF (\textcolor{blue}{UAI2009}) & 0.0876 &  0.0749 & 0.1503 &  0.0966 & 0.0437 & 0.0264 & 0.0689 & 0.0339 & 0.0341 & 0.0210 & 0.0560 & 0.0276 \\
    & LightGCN (\textcolor{blue}{SIGIR2020}) & 0.0987 &  0.0833 &  0.1707 & 0.1083& 0.0534 & 0.0325  & 0.0822 & 0.0411  & 0.0540 & 0.0325 & 0.0904 & 0.0436 \\
    \hline
    & CDAE (\textcolor{blue}{WSDM2016}) & 0.0991 &  0.0829 & 0.1705 & 0.1078 & 0.0538 & 0.0361 & 0.0737 & 0.0422 & 0.0444 & 0.0280 & 0.0703 & 0.0360 \\
    
    & MultiDAE (\textcolor{blue}{WWW2018}) & 0.0995 &  0.0803 &   0.1753 &  0.1067& 0.0571 & 0.0357 & 0.0855 & 0.0422 & 0.0522 & 0.0316 & 0.0864 & 0.0419 \\
    \hline
    & HyperML (\textcolor{blue}{WSDM2020}) & 0.0997 & 0.0832 & 0.1752 & 0.1042 & 0.0567 & 0.0362 & 0.0846 &  0.0432 & 0.0539 & 0.0311 & 0.0911 & 0.0409\\
    & HGCF(\textcolor{blue}{WWW2021}) & 0.1009 & 0.0865 & 0.1771 & 0.1126& 0.0633 & 0.0392 & 0.0931 & 0.0481  & 0.0560 & 0.0329 & 0.0931 & 0.0447 \\
      
    & HICF (\textcolor{blue}{KDD2022}) & 0.0970 & 0.0848  & 0.1754 & 0.1010 & 0.0652 & 0.0426 & 0.0984 & 0.0514 & \underline{0.0590} & \underline{0.0366} & \underline{0.0968} & \underline{0.0488}\\
     \hline
    & CODIGEM (\textcolor{blue}{KSEM2022}) & 0.0972 &  0.0837 & 0.1699 & 0.1087 & 0.0300 & 0.0192 & 0.0478 & 0.0245 & 0.0470 & 0.0292 & 0.0775 & 0.0385 \\

    & DiffRec (\textcolor{blue}{SIGIR2023}) & \underline{0.1023} & \underline{0.0876} &   \underline{0.1778} &  \underline{0.1136} & \underline{0.0695} & \underline{0.0451} & \underline{0.1010} & \underline{0.0547} & 0.0581 & 0.0363 & 0.0960 & 0.0478 \\
    & DDRM (\textcolor{blue}{SIGIR2024}) & 0.1017 &  0.0874 & 0.1760 & 0.1132 & 0.0685 & 0.0432 & 0.0994 & 0.0521 & 0.0556 & 0.0343 & 0.0943 & 0.0438 \\
    \hline

    \rowcolor{gray!20} &{HDRM}  & \textbf{0.1078*} & \textbf{0.0931*} & \textbf{0.1852*} & \textbf{0.1190*}  & \textbf{0.0698*} & \textbf{0.0457*} & \textbf{0.1057*} & \textbf{0.0582*} &  \textbf{0.0623*} &   \textbf{0.0390*} & \textbf{0.1024*} & \textbf{0.0499*}  \\
    \hline
    \hline
    &Improv. & 5.4\% & 6.3\% & 4.2\% & 4.8\% & 0.5\% & 1.3\% & 4.7\% & 6.5\% & 5.6\% & 6.7\% & 5.8\% & 2.4\%\\
    \bottomrule
    \end{tabular}}
    \label{tab2}
\end{table*}
\subsubsection{Baselines and Hyper-parameter Settings} 

The effectiveness of our method is assessed through comparison with the following baselines: classic collaborative filtering methods include BPRMF ~\cite{rendle2009bpr} and LightGCN~\cite{he2020lightgcn}. Autoencoder-based recommender methods are represented by CDAE ~\cite{wu2016collaborative} and Multi-DAE ~\cite{liang2018variational}. Diffusion-based recommender methods include CODIGEM~\cite{walker2022recommendation}, DiffRec~\cite{li2023diffurec}, and DDRM~\cite{zhao2024denoising}. Finally, hyperbolic recommender methods encompass HyperML ~\cite{vinh2020hyperml}, HGCF ~\cite{sun2021hgcf}, and HICF ~\cite{yang2022hicf}. It is worth noting that the complete form of our adopted DDRM is LightGCN+DDRM. Further details on these models can be found in Appendix~\ref{baselines}. More details about our HDRM's hyper-parameter settings can be found in Appendix~\ref{hyperparameter}.


\begin{table*}
\centering
\small 
\caption{Performance of different design variations on the three datasets. The bolded numbers denote the most significant change in performance.}
 \renewcommand{\arraystretch}{1.1}
 \setlength{\tabcolsep}{1.8mm}{
    \begin{tabular}{lcccccccccccccl}
     \toprule
     & \multirow{2}{*}{Model}& \multicolumn{4}{c}{ML-1M}&\multicolumn{4}{c}{Amazon-Book} &\multicolumn{4}{c}{Yelp2020}
     \\
    \cmidrule(r){3-6} \cmidrule(r){7-10} \cmidrule(r){11-14}
    & & R@10 & N@10 &R@20 &N@20 & R@10 & N@10 &R@20 & N@20 & R@10 & N@10 &R@20  &N@20 \\
    \hline
    \hline
& HDRM & 0.1078 & 0.0931 & 0.1852 & 0.1190 & 0.0698 & 0.0467 & 0.1057 & 0.0582 & 0.0623 & 0.0390 & 0.1024 & 0.0499\\
\hline
    & HDRM w/o $\mathcal{H}^n_\kappa$ & 0.1063 & 0.0917  & 0.1793 & 0.1151 & 0.0695 & 0.0447 & 0.1025 & 0.0561 & 0.0589 & 0.0377 & 0.0986 & 0.0473\\
    & HDRM w/o Geo & 0.1052 & 0.0902 &   0.1778 & 0.1136 & \textbf{0.0687} & \textbf{0.0438} & \textbf{0.1001} & \textbf{0.0528} & \textbf{0.0571} & \textbf{0.0362} & \textbf{0.0958} & \textbf{0.0453} \\
    & HDRM w/o Diff & \textbf{0.1035} & \textbf{0.0883} &   \textbf{0.1763} &  \textbf{0.1131} & 0.0693 & 0.0446 & 0.1008 & 0.0541 & 0.0587 & 0.0373 & 0.0967 & 0.0467 \\
    \bottomrule
    \end{tabular}}
    \label{tab_abl}
\end{table*}
\subsection{Overall Performance Comparison (RQ1)}
Table \ref{tab2} reports the comprehensive performance of all the compared baselines across three datasets. Based on the results, the  main observations are as follow:

\begin{itemize}[leftmargin=*]
    \item Our proposed HDRM demonstrates consistent performance improvements across all metrics on three datasets compared to state-of-the-art baselines. This superior performance is primarily attributed to three key factors: 1) HDRM excels in capturing the complex relationships in user-item interactions compared to Euclidean-based approaches. This capability allows for a more nuanced understanding of the underlying recommendation dynamics. 2) By employing neural networks to incrementally learn each denoising transition step from $t$ to $t$-1, HDRM effectively models complex distributions. This approach significantly enhances the model's capacity to capture intricate patterns in the data. 3) Through learning the data distribution, HDRM exhibits superior capabilities in addressing data sparsity issues. This enables the model to infer latent associations from limited data.
    \item Diffusion-based approaches, such as DDRM and DiffRec, generally outperform traditional methods like BPRMF and LightGCN. This superior performance can be attributed to the alignment between their generative frameworks and the processes underlying user-item interactions. Among the generative methods, DiffRec demonstrates particularly impressive results, leveraging variational inference and KL divergence to achieve more robust generative modeling. In contrast, CODIGEM underperforms compared to LightGCN and other generative methods, primarily due to its reliance on only the first autoencoder for inference.
    \item Diffusion-based recommendation models do not universally outperform hyperbolic-based models. For instance, on the Yelp2020 dataset, HICF demonstrates superior performance compared to DiffRec. While diffusion-based models exhibit enhanced robustness and noise-handling capabilities, hyperbolic spaces are inherently well-suited for representing data with hierarchical structures and power-law distributions—characteristics that closely align with user-item interaction graphs in numerous recommender systems. Notably, models that integrate hyperbolic geometry with diffusion techniques have exhibited superior performance across three datasets by leveraging the strengths of both approaches.
\end{itemize}

\begin{table*}
\centering
\small 
\caption{Comparative analysis of best diffusion methods (DiffRec) and hyperbolic approaches (HICF) in noisy datasets, focusing on their performance amid random clicks and other data imperfections, highlighting the best and second-best performances in bold and borderline, respectively. Numbers with an
asterisk (*) indicate statistically significant improvements over the best baseline (t-test with p-value <0.05).}
 \renewcommand{\arraystretch}{1.1}
 \setlength{\tabcolsep}{1.mm}{
    \begin{tabular}{lcccccccccccccl}
     \toprule
     & \multirow{2}{*}{Model}& \multicolumn{4}{c}{ML-1M}&\multicolumn{4}{c}{Amazon-Book} &\multicolumn{4}{c}{Yelp2020}
     \\
    \cmidrule(r){3-6} \cmidrule(r){7-10} \cmidrule(r){11-14}
    & & R@10 & N@10 &R@20 &N@20 & R@10 & N@10 &R@20 & N@20 & R@10 & N@10 &R@20  &N@20 \\
    \hline
    & HICF (\textcolor{blue}{KDD2022}) & 0.0635 & 0.0437  & 0.1211 & 0.0643 & 0.0512 & 0.0298 & 0.0763 & 0.0374 & 0.4770 & 0.0286 & 0.815 & 0.0387\\

    & DiffRec (\textcolor{blue}{SIGIR2023}) & 0.0658 & 0.0488 &   \underline{0.1236} & 0.0703 & \underline{0.0537} & \underline{0.0329} & \underline{0.0806} & \underline{0.0411} & 0.0501 & \underline{0.0307} & 0.0847 & \underline{0.0412} \\

    & DDRM (\textcolor{blue}{SIGIR2024}) & \underline{0.0667} & \underline{0.0508} &   0.1221 &  \underline{0.0710} & 0.0468 & 0.0273 & 0.0742 & 0.0355 & \underline{0.0516} & 0.0305 & \underline{0.0870} & \underline{0.0412} \\

    \rowcolor{gray!20} &{HDRM}  & \textbf{0.0679*} & \textbf{0.0522*} & \textbf{0.1254*} & \textbf{0.0714*}  & \textbf{0.0554*} & \textbf{0.0336*} & \textbf{0.0819*} & \textbf{0.0427*} &  \textbf{0.0523*} &   \textbf{0.0325*} & \textbf{0.0883*} & \textbf{0.0432*}  \\
    \bottomrule
    \end{tabular}}
    \label{tab3}
\end{table*}

\subsection{Ablation Study (RQ2)}
To validate the effectiveness of our proposed method, we conducted ablation studies by removing three key components from HDRM: the hyperbolic encoder (HDRM w/o $\mathcal{H}^n_\kappa$), geometric restrictions (HDRM w/o Geo) and diffusion model (HDRM w/o Diff). Table \ref{tab_abl} presents the results of our experiments on three datasets, from which we draw the following significant conclusions:

\begin{itemize}[leftmargin=*]
    \item The model's performance significantly decreases when the diffusion model, geometric restrictions, and hyperbolic encoder are removed individually. This demonstrates the crucial role these modules play in the model's effectiveness. Furthermore, the table \ref{tab_abl} reveals that the absence of the diffusion model and geometric restrictions has a more substantial impact on the model's performance compared to the hyperbolic encoder. This discrepancy may be attributed to the inherent hierarchical structure and information-rich properties of hyperbolic space. However, without geometric restrictions, the learned embeddings might become overly dispersed or concentrated within the space, failing to fully leverage the advantages of hyperbolic geometry. In contrast to the diffusion component, real-world recommendation models may rely more heavily on capturing the propagation and evolution of preferences rather than strictly adhering to hierarchical structures.
    \item The removal of the diffusion model results in the most significant performance decline on the ML-1M dataset, while the elimination of geometric restrictions leads to the most substantial performance drop on the Amazon-Book and Yelp2020 datasets. This discrepancy may be attributed to the higher density of the ML-1M dataset compared to Amazon-Book and Yelp2020. The marked performance degradation observed when removing the diffusion model from the relatively dense ML-1M dataset underscores the critical role of the diffusion process in modeling complex and dynamic user behaviors. The higher density of ML-1M implies more frequent user-item interactions and intricate information flow compared to other datasets. In such an environment, diffusion models may more effectively capture rapidly evolving user preferences, social influences, and non-linear relationships.
\end{itemize}

In conclusion, our ablation studies highlight the significant contributions of each module in HDRM to the overall model performance. These findings not only validate our design choices but also provide insights into the relative importance of different components in hyperbolic recommender models.

\begin{figure*}[t]
	 \centering
	\begin{minipage}{0.32\linewidth}
		\vspace{10pt}
\centerline{\includegraphics[width=1.0\textwidth]{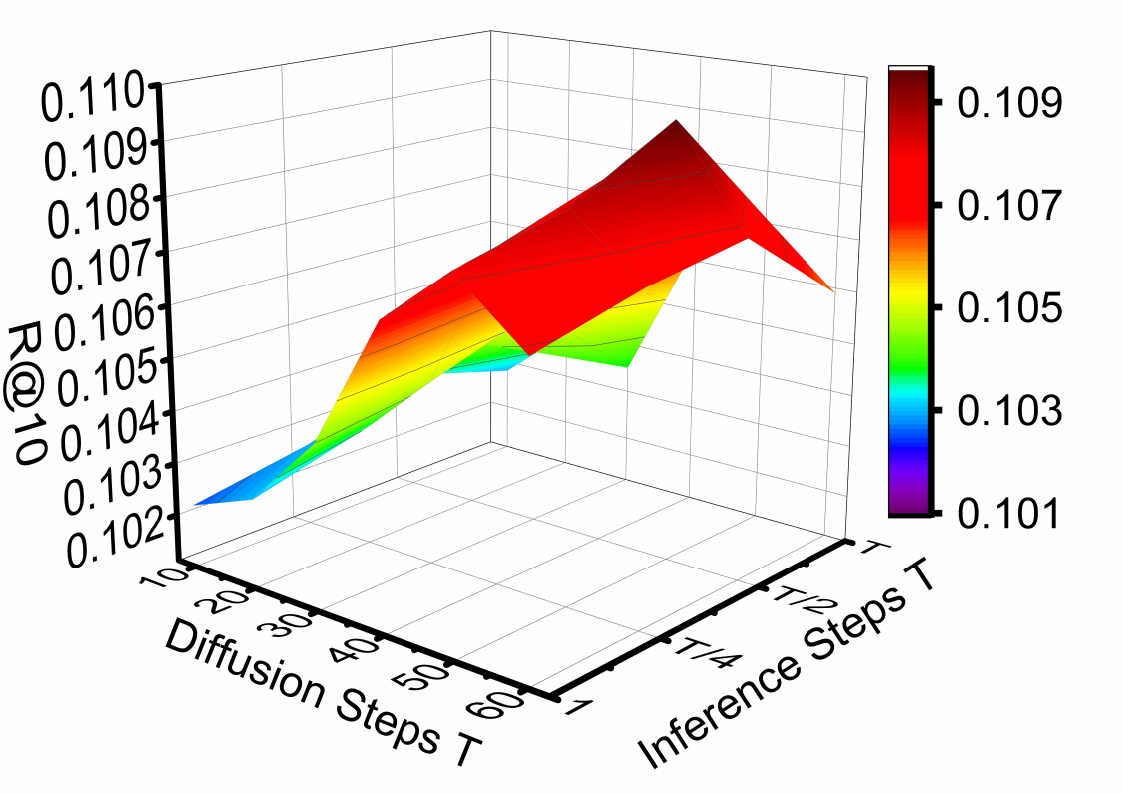}}
		\centerline{(a) ML-1M}
	\end{minipage}
 \begin{minipage}{0.32\linewidth}
		\vspace{10pt}
\centerline{\includegraphics[width=1.0\textwidth]{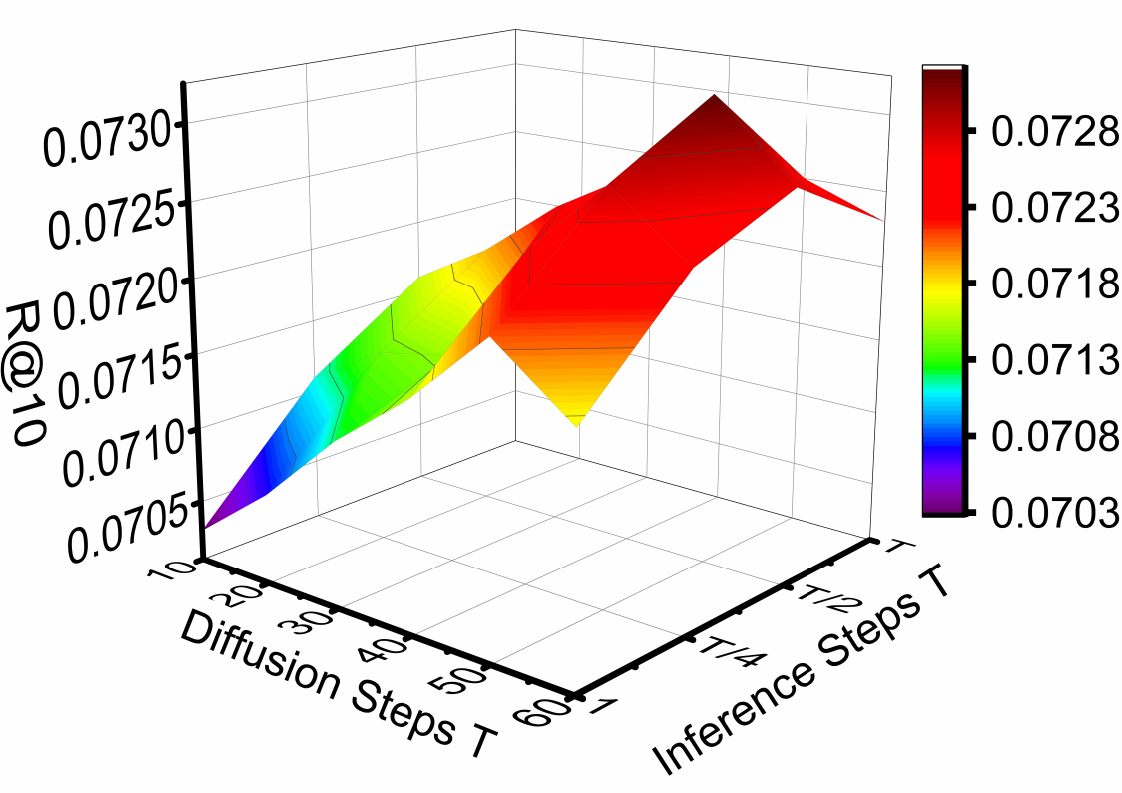}}
		\centerline{(b) Amazon-Book}
	\end{minipage}
 \begin{minipage}{0.32\linewidth}
        \vspace{10pt}
\centerline{\includegraphics[width=1.\textwidth]{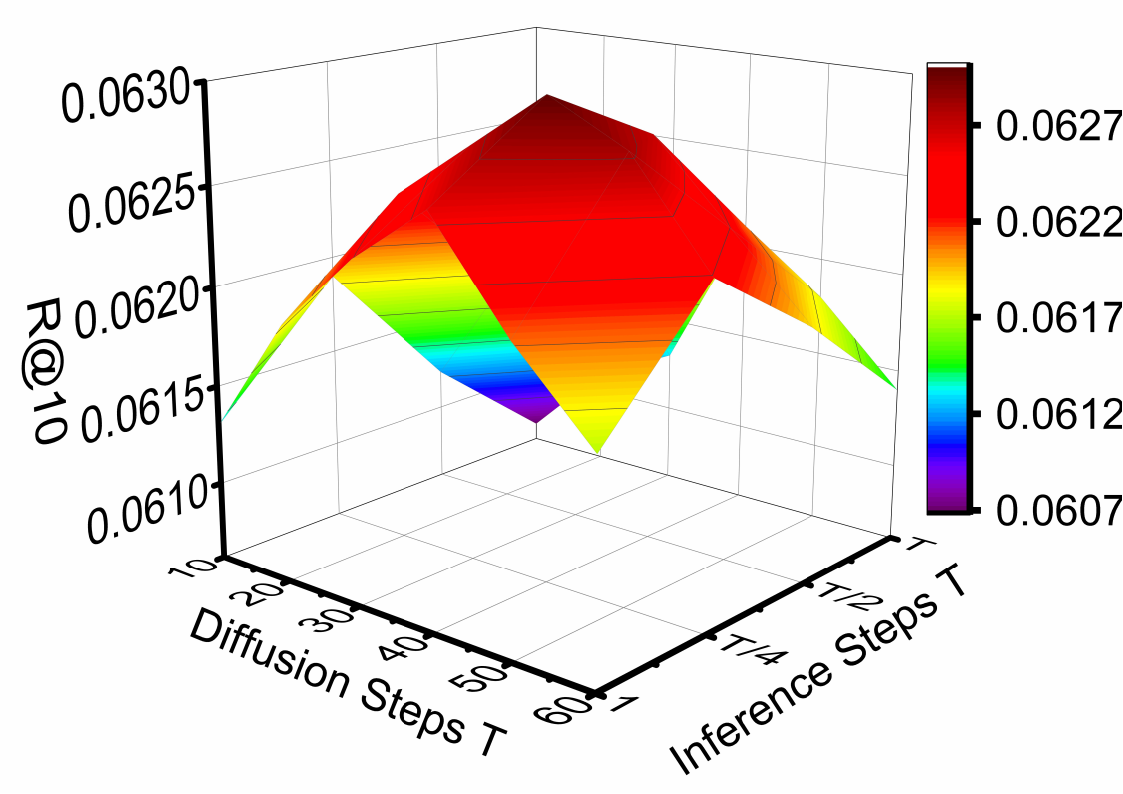}}
        \centerline{(c) Yelp2020}
    \end{minipage}
   \caption{The variation of model performance (R@10) across three datasets as diffusion steps and inference steps change.}
	\label{fig_step}
\end{figure*}
\subsection{Robustness Analysis (RQ3)}
In real-world recommender systems, user behavior data often contains noise, such as random clicks or unintentional interactions. To evaluate HDRM's effectiveness in handling noisy data, we conducted a comparative analysis with DiffRec and DDRM, the leading diffusion methods, and HICF, the leading hyperbolic approach. Our noise comprises natural noise (\textit{cf.} Table \ref{tab01}) and randomly sampled interactions, maintaining an equal scale for both components. 

Table \ref{tab3} presents the performance metrics of these models in the presence of noise. The results demonstrate that HDRM consistently outperforms both HICF, DDRM and DiffRec, validating its robustness against noisy data. Notably, diffusion-based models exhibit superior performance in noisy environments, which aligns with theoretical expectations. This can be attributed to the inherent denoising process that underpins diffusion models, making them particularly well-suited for mitigating the impact of erroneous user interactions. In contrast, HICF's performance degraded significantly in the presence of noise, suggesting that the hyperbolic space does not offer a substantial advantage over Euclidean space in terms of reducing the influence of noisy interactions. This finding challenges the presumed benefits of hyperbolic embeddings in this context and highlights the need for further investigation into their limitations in noisy recommendation scenarios.

\subsection{In-depth Analysis (RQ4)}
\subsubsection{Diffusion Step Analysis}
We investigate the impact of varying diffusion and inference steps on HDRM's performance. Figure \ref{fig_step} illustrates our experimental results across three datasets, HDRM's performance initially improves as diffusion and inference steps increase. However, it subsequently declines with further increases in these steps. This phenomenon can be attributed to several factors. When the number of diffusion steps is insufficient, the model lacks adequate iterations to progressively refine recommendation results, leading to suboptimal capture of user preferences. Conversely, an excessive number of diffusion steps may cause the model to overfit the noise distribution, potentially discarding valuable information from the original data. Similarly, an insufficient number of inference steps prevents the model from fully recovering the original data distribution from a pure noise state. However, an excessive number of inference steps can result in over-optimization, potentially causing the model to deviate from the target distribution. More diffusion step results can be found in Appendix \ref{n10}.

\begin{figure}[t]
	\centering
	\setlength{\fboxrule}{0.pt}
	\setlength{\fboxsep}{0.pt}
	\fbox{      \includegraphics[width=1.0\linewidth]{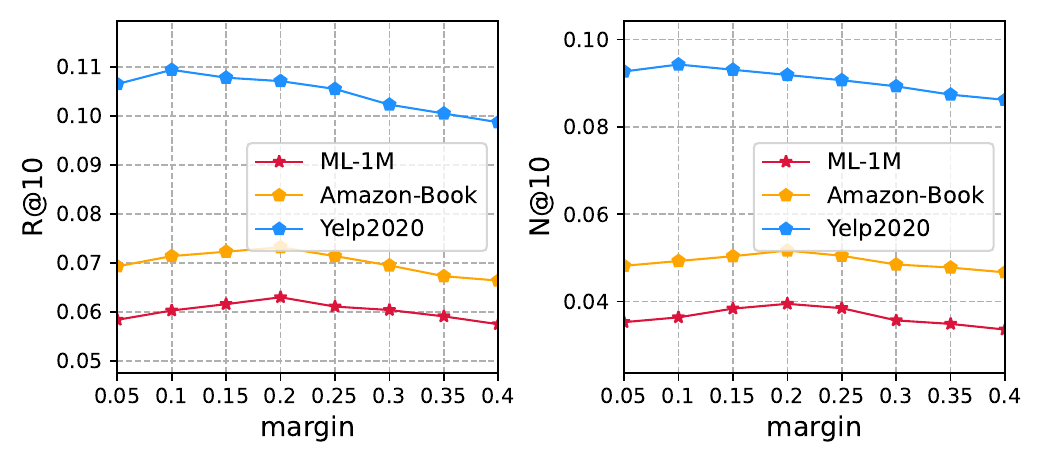}
	}
	\caption{The variation of model performance across three datasets as the margin changes.
 }
	\label{pic_margin}
\end{figure}
\subsubsection{Margin Analysis}
\label{margin_analysis}
We investigate the impact of varying margin values on HDRM's performance. Figure \ref{pic_margin} presents the experimental results, revealing a non-monotonic relationship between margin size and HDRM's performance. As the margin increases, HDRM's performance initially improves before subsequently declining, indicating the existence of an optimal margin value for maximizing model effectiveness. On the ML-1M dataset, the model achieves peak performance at a margin of 0.1. In contrast, for the Amazon-Book and Yelp2020 datasets, optimal model performance is attained at a margin of 0.2. This discrepancy is notable across different datasets. The Amazon-Book and Yelp2020 datasets show greater distinction between positive and negative samples than ML-1M. Considering the hyperbolic margin loss function, the margin represents the expected difference in scores between positive and negative samples. When dealing with datasets characterized by substantial disparities between positive and negative samples, a larger margin is advisable.

\section{RELATED WORK}
In this section, we review two relevant prior works: hyperbolic representation learning and generative recommendation.

\subsection{Hyperbolic Representation Learning}
Currently, Non-Euclidean representation learning, particularly hyperbolic representation learning, plays a crucial role in recommender systems (RSs)\cite{vinh2020hyperml, sun2021hgcf, zhang2022geometric}. HyperML\cite{vinh2020hyperml} investigates metric learning in hyperbolic space and its connection to collaborative filtering. Similarly, HGCF~\cite{sun2021hgcf} proposes a hyperbolic GCN model for CF. In order to address the power-law distribution in recommender systems, HICF~\cite{yang2022hicf} focuses on enhancing the attention towards tail items in hyperbolic spaces, incorporating geometric awareness into the pull and push process. Interestingly, GDCF~\cite{zhang2022geometric} aims to capture intent factors across geometric spaces by learning geometric disentangled representations associated with user intentions and different geometries. On the other hand,
paper~\cite{xu2020learning} highlights that the naive inner product used in the factorization machine model~\cite{rendle2010factorization} may not adequately capture spurious or implicit feature interactions. Collaborative metric learning~\cite{hsieh2017collaborative} proposes that learning the distance instead of relying on the inner product provides benefits in capturing detailed embedding spaces that encompass item-user interactions, item-item relationships, and user-user distances simultaneously. Consequently, the triangle inequality emerges as a more favorable alternative to the inner product. Paper~\cite{lin2023hyperbolic} is the first to introduce the use of hyperbolic diffusion geometry to reveal hierarchical structures. Similarly, HypDiff~\cite{fu2024hyperbolic} leverages hyperbolic diffusion for graph generation.

Inspired by previous works on hyperbolic models, HDRM builds upon these foundations to address challenges in recommendation systems. Unlike existing methods, HDRM is specifically designed for this context, leveraging the geometric properties of hyperbolic space, particularly its anisotropy, to guide the diffusion of user preferences. By aligning the diffusion process with the underlying structure of user interests, HDRM effectively models preferences and captures the complexities of the recommendation task.
\vspace{-1mm}
\subsection{Generative Recommendation}
Generative models, such as Generative Adversarial Networks (GANs) \cite{gao2021recommender, jin2020sampling, wang2017irgan} and Variational Autoencoders (VAEs) \cite{liang2018variational, ma2019learning, zhang2017autosvd++}, play an important role in personalized recommendations but suffer from structural drawbacks \cite{kingma2016improved, sohl2015deep}. Recently, diffusion models have emerged as an alternative, offering better stability and representation capabilities, especially in recommendation systems \cite{chen2024adversarial, jiang2024diffkg, ma2024plug, wu2020diffnet++, wu2019neural}. Models like CODIGEM \cite{walker2022recommendation} and DiffRec \cite{wang2023diffusion} use diffusion models to predict user preferences by simulating interaction probabilities. Meanwhile, other approaches \cite{du2023sequential, li2023diffurec, liu2023diffusion, wang2024conditional, zhao2024denoising} focus on content generation at the embedding level. For instance, DiffRec \cite{li2023diffurec} and CDDRec \cite{wang2024conditional} add noise to target items in the forward process, later reconstructing them based on users' past interactions. DiffuASR \cite{liu2023diffusion} applies diffusion models to generate item sequences, addressing data sparsity challenges. Furthermore, DDRM~\cite{zhao2024denoising} leverages diffusion models to denoise implicit feedback, leading to more robust representations in learning tasks. 

Unlike the above diffusion models that address data noise in recommender systems, HDRM emphasizes the underlying structure of item data by designing a directional diffusion process that more closely aligns with the data’s inherent characteristics, thereby preserving the structural properties of the original distribution.

\section{CONCLUSION}
Motivated by the promising results obtained from recent diffusion-based recommender models~\cite{wang2023diffusion, zhao2024denoising, li2024multi}, we have decided to explore a more complex geometry architecture. Building on the success of hyperbolic representation learning
methods~\cite{lin2023hyperbolic, sun2021hgcf, yang2022hicf, fu2024hyperbolic}, we investigate that they hold great potential in addressing the non-Euclidean structural anisotropy of the underlying diffusion process in user-item interaction graphs. To this end, 
we propose HDRM model architecture, further experiments demonstrate the superiority of this method. We believe that this paper represents a milestone in hyperbolic diffusion models and offers a valuable baseline for future research in this field.

\begin{acks}
This research work is supported by the National Key Research and Development Program of China under Grant No. 2021ZD0113602, the National Natural Science Foundation of China under Grant Nos. 62176014, 62276015, the Fundamental Research Funds for the Central Universities.
\end{acks}

\bibliographystyle{ACM-Reference-Format}
\bibliography{reference}
\clearpage

\appendix

\section{Methods}
\begin{table*}[h]
    \centering
    \small 
    \caption{Summary of operations in the Poincaré ball manifold and the Lorentz manifold}

    \renewcommand{\arraystretch}{1.2} 
     \setlength{\tabcolsep}{0.4mm}{ 
    \begin{tabular}{lcl}
     \toprule
        & \textbf{Poincaré Ball Manifold} & \textbf{Lorentz Manifold} \\ \hline \hline
        \textbf{Notation} & 
        $\mathcal{B}^n_\kappa = \left\{ \mathbf{x} \in \mathbb{R}^n : \langle \mathbf{x}, \mathbf{x} \rangle_2 < -\frac{1}{\kappa} \right\}$ &  
        $\mathcal{L}^n_\kappa = \left\{ \mathbf{x} \in \mathbb{R}^{n+1} : \langle \mathbf{x}, \mathbf{x} \rangle_{\mathcal{L}} = \frac{1}{\kappa} \right\}$ \\ 
        \textbf{Geodesics distance} & 
        $d_{\mathcal{B}}^\kappa(\mathbf{x}, \mathbf{y}) = \frac{1}{\sqrt{|\kappa|}} \cosh^{-1} \left( 1 - \frac{2\kappa \|\mathbf{x} - \mathbf{y}\|_2^2}{(1+\kappa\|\mathbf{x}\|_2^2)(1+\kappa\|\mathbf{y}\|_2^2)} \right)$ & 
        $d_{\mathcal{L}}^\kappa(\mathbf{x}, \mathbf{y}) = \frac{1}{\sqrt{|\kappa|}} \cosh^{-1} (\kappa \langle \mathbf{x}, \mathbf{y} \rangle_{\mathcal{L}})$ \\ 
        \textbf{Logarithmic map} & 
        $\log_{\mathbf{x}}^\kappa (\mathbf{y}) = \frac{2}{\sqrt{|\kappa|} \lambda_{\mathbf{x}}^\kappa} \tanh^{-1} \left( \sqrt{|\kappa|} \| - \mathbf{x} \oplus_\kappa \mathbf{y} \|_2 \right) \frac{- \mathbf{x} \oplus_\kappa \mathbf{y}}{\| - \mathbf{x} \oplus_\kappa \mathbf{y} \|_2}$ & 
        $\log_{\mathbf{x}}^\kappa (\mathbf{y}) = \frac{\cosh^{-1} (\langle \mathbf{x}, \mathbf{y} \rangle_{\mathcal{L}})}{\sinh \left( \cosh^{-1} (\kappa \langle \mathbf{x}, \mathbf{y} \rangle_{\mathcal{L}}) \right)} (\mathbf{y} - \kappa \langle \mathbf{x}, \mathbf{y} \rangle_{\mathcal{L}} \mathbf{x})$ \\ 
        \textbf{Exponential map} & 
        $\exp_{\mathbf{x}}^\kappa (\mathbf{v}) = \mathbf{x} \oplus_\kappa \left( \tanh \left( \sqrt{|\kappa|} \frac{\lambda_{\mathbf{x}}^\kappa \|\mathbf{v}\|_2}{2} \right) \frac{\mathbf{v}}{\sqrt{|\kappa|} \|\mathbf{v}\|_2} \right)$ & 
        $\exp_{\mathbf{x}}^\kappa (\mathbf{v}) = \cosh \left( \sqrt{|\kappa|} \|\mathbf{v}\|_{\mathcal{L}} \right) \mathbf{x}$ \\ 
        \textbf{Parallel transport} & 
        $P\mathcal{T}_{\mathbf{x} \to \mathbf{y}}^\kappa (\mathbf{v}) = \frac{\lambda_{\mathbf{x}}^\kappa}{\lambda_{\mathbf{y}}^\kappa} \text{gyr} [\mathbf{y}, -\mathbf{x}] \mathbf{v}$ &  
        $P\mathcal{T}_{\mathbf{x} \to \mathbf{y}}^\kappa (\mathbf{v}) = \mathbf{v} - \frac{\kappa \langle \mathbf{y}, \mathbf{v} \rangle_{\mathcal{L}}}{1+\kappa \langle \mathbf{x}, \mathbf{y} \rangle_{\mathcal{L}}} (\mathbf{x} + \mathbf{y})$ \\ 
        \textbf{Origin point} & 
        $\mathbf{0}_n$ & \qquad \qquad \quad
        $\left[ \frac{1}{\sqrt{|\kappa|}}, \mathbf{0}_n \right]$ \\ \bottomrule
    \end{tabular}}
    \label{tab0}
\end{table*}

\label{sec:partone}
\subsection{Hyperbolic Spaces}
\label{Hyperbolic Spaces}
Here, we provide a comparison of geometric operations~\cite{yang2023kappahgcn} between the Poincaré ball manifold and the Lorentz manifold as summarized in Table~\ref{tab0}. It outlines the notation, geodesic distance, logarithmic map, exponential map, parallel transport, and the origin point for both manifolds, along with their respective mathematical formulations. This table serves to summarize the computational methods for these operations across the two different manifolds, highlighting their similarities and differences.

\subsection{Further Exploration of Hyperbolic Clustering}
\label{proof_hyperbolic_clustering}

In this section, we further explore certain phenomena of hyperbolic clustering, particularly in the context of its prominent hierarchical structure.

\subsubsection{Hyperbolic Embeddings}
\label{sec:hyperbolic_embeddings}
Here, we discuss the concept of embedding data points into hyperbolic space, particularly within the Lorentz manifold, and highlight the key geometric properties that facilitate clustering. 

In hyperbolic clustering, the objective is to embed a set of entities \( \{ \mathbf{x}_1, \mathbf{x}_2, \dots, \mathbf{x}_N \} \) into the Poincaré ball model, ensuring that the relationships defined by the similarity between the items are preserved. As shown in Table~\ref{tab0}, Lorentz manifold is denoted as:
\begin{equation}
\mathcal{L}^n_\kappa = \left\{ \mathbf{x} \in \mathbb{R}^{n+1} : \langle \mathbf{x}, \mathbf{x} \rangle_\mathcal{L} = \frac{1}{\kappa} \right\},
\end{equation}
where \(\langle \mathbf{x}, \mathbf{x} \rangle_2\) represents the Lorentz product of the point \(\mathbf{x}\) with itself, and \(\kappa\) is the curvature of the space. Next, the hyperbolic distance between two points \(\mathbf{x}_i\) and \(\mathbf{x}_j\) in the Lorentz manifold is defined as:
\begin{equation}
d_{\mathcal{L}}^\kappa(\mathbf{x}, \mathbf{y}) = \frac{1}{\sqrt{|\kappa|}} \cosh^{-1} (\kappa \langle \mathbf{x}, \mathbf{y} \rangle_{\mathcal{L}}).
\end{equation}

The geodesic distance formula effectively captures the inherent hierarchical structure within the data. By utilizing this distance metric, data points can be embedded into hyperbolic space, where hierarchical relationships are preserved more naturally than in Euclidean space.

\subsubsection{Hierarchical Clustering in Hyperbolic Space}
\label{sec:hyperbolic_hierarchical_clustering}

Conventional Euclidean clustering algorithms, such as agglomerative clustering or k-means, can still be applied as shown in Eq.\eqref{hyper-kmeans}, the key distinction lies in replacing the traditional Euclidean distance metric with the hyperbolic distance.

The next step in hierarchical clustering with hyperbolic embeddings is to identify the Lowest Common Ancestor (LCA) of two embeddings. In hyperbolic geometry, the LCA of two embeddings \( \mathbf{x}_i \) and \( \mathbf{x}_j \) is defined as the point along their geodesic path that is closest to the origin of the manifold. Mathematically, this can be expressed as:
\begin{equation}
\mathbf{x}_i \vee \mathbf{x}_j = \arg \min_{\mathbf{x}_o \in \mathcal{L}_\kappa} d_\mathcal{L}^{\kappa}(\mathbf{o}, \mathbf{x}_o),
\end{equation}
where \( \mathbf{o} \) denotes the anchor of the manifold. Intuitively, the LCA provides a natural hierarchical relationship between the two embeddings by identifying the closest point to the origin along their connecting geodesic. Functionally, the LCA in hyperbolic space, analogous to its counterpart in discrete tree structures, identifies the closest common point along the geodesic path between two embeddings, capturing their hierarchical relationship.

\subsubsection{Optimization of Hyperbolic Embeddings}
\label{sec:hyperbolic_optimization}

The optimization of hyperbolic embeddings is a key component in the hierarchical clustering process. The goal is to optimize the hyperbolic embeddings \( \mathbf{X} \) is denoted as $\mathbf{X} = \left\{ \mathbf{x}_1, \mathbf{x}_2, \dots, \mathbf{x}_N \right\}$, such that the embeddings of similar items are placed closer together in hyperbolic space, while preserving the hierarchical relationships inherent in the data. To achieve this, the following objective function is minimized, based on the similarity \( \mathbf{S} \) and the hyperbolic distance metric \( d_\mathcal{L}(\mathbf{x}_i, \mathbf{x}_j) \):
\begin{equation}
\min_{\mathbf{X}} \sum_{i,j} \mathbf{S}_{ij} \cdot d_\mathcal{L}(\mathbf{x}_i, \mathbf{x}_j),\;\text{where} \;\mathbf{S}_{ij} = \langle \log_{\mathbf{x}_i}^{\kappa}(\mathbf{x}_j), \log_{\mathbf{x}_i}^{\kappa}(\mathbf{x}_i) \rangle.
\end{equation}

This optimization computes the inner product of the embeddings in the tangent space, utilizing the Euclidean geometry of the space. By minimizing this objective function, hyperbolic embeddings are obtained that both preserve the hierarchical structure of the data and remain within the boundary of the Lorentz manifold. Functionally, this process ensures that hierarchical relationships are maintained within the non-Euclidean space of hyperbolic geometry, facilitating more accurate clustering and representation of complex structures.

\subsection{Discussion on the Poincaré Normal Distribution}
\label{proof_poincare_normal_distribution}

\subsubsection{Poincaré Normal Distribution and Its Definition}
\label{poincare_normal_dist}

In the forward diffusion process, we assume that the noise \( \epsilon_\mathcal{B} \) follows a Poincaré normal distribution:
\begin{equation}
\epsilon_\mathcal{B} \sim \mathcal{N}_\mathcal{B}(0, \mathrm{I}),
\end{equation}
where \( \mathcal{N}_\mathcal{B}(0, \mathrm{I}) \) represents the Poincaré normal distribution with mean \( 0 \) and covariance matrix \( \mathrm{I} \) (the identity matrix). This distribution signifies that the noise is isotropic, with unit variance along each dimension. The probability density function of this distribution is given by:
\begin{equation}
f(\mathbf{\epsilon}_\mathcal{B}) = \sqrt{\frac{2}{\pi}} e^{-\frac{|\mathbf{\epsilon}_\mathcal{B}|^2}{2}},
\end{equation}
where \( |\mathbf{\epsilon}_\mathcal{B}|^2 \) is the squared Euclidean norm of the noise vector \( \mathbf{\epsilon} \). When \( \mu = 0 \), the distribution simplifies to the half-normal distribution. This distribution plays a key role in describing the noise dynamics during the diffusion process in non-Euclidean spaces.

\subsubsection{Impact of Poincaré Noise on Diffusion Process}
\label{impact_poincare_noise}

In the context of the forward diffusion process, the noise \( \epsilon_\mathcal{B} \) impacts the system at each diffusion step, allowing us to capture the anisotropic structural features of the underlying space. As the diffusion process progresses over time, the system’s states are perturbed by this noise, which evolves in both mean and variance over time.

The evolution of the state \( \mathbf{z}_t \) over time can be modeled as:
\begin{equation}
\begin{split}
\mathbf{z}_t = \eta \cdot \mathbf{z}_{t-1} + \epsilon_\mathcal{B}, \quad \epsilon_\mathcal{B} \sim \mathcal{N}_\mathcal{B}(0, \mathrm{I}) \\
\implies
\lim_{t \to \infty} \mathbf{z}_t \sim \mathcal{N}_\mathcal{B} (\delta \mathbf{z}_{t-1}, \mathrm{I}).
\end{split}
\end{equation}
Here, \( \eta \) represents the scaling factor, and \( \mathbf{z}_t \) is the state of the system at time \( t \). The term \( \epsilon_\mathcal{B} \) denotes the Poincaré noise, and it follows the distribution \( \mathcal{N}_\mathcal{B}(0, \mathrm{I}) \).

As the diffusion process continues, the state \( \mathbf{z}_t \) evolves according to the noise dynamics, leading to a final state distribution that is governed by the Poincaré normal distribution with mean \( \delta \mathbf{z}_{t-1} \) and covariance matrix \( \mathrm{I} \).

The long-term behavior of the diffusion process can be described by the following probability distribution:
\begin{equation}
p(\mathbf{z}_t | \mathbf{z}_0) = \mathcal{N}_\mathcal{B}(\mu_t, \sigma_t),
\end{equation}
where \( \mu_t \) and \( \sigma_t \) are defined as:
\begin{equation}
\mu_t = \sqrt{\overline{\alpha}_t} + \delta \tanh \left( \frac{\sqrt{\kappa} \lambda_o^\kappa(t)}{T_0} \right),\;\;\\
\sigma_t = (1 - \overline{\alpha_t}) \mathrm{I}.
\end{equation}

As \( t \to \infty \), the distribution of the system's state converges to:
\begin{equation}
\lim_{t \to \infty} \mathbf{z}_t \sim \mathcal{N}_\mathcal{B}(\delta \mathbf{z}_0, \mathrm{I}).
\end{equation}

This result demonstrates the long-term behavior of the diffusion process, where the mean shifts based on the initial state \( \mathbf{z}_0 \), and the variance remains constant at \( \mathrm{I} \). The Poincaré normal distribution is critical for capturing the complex geometry of the diffusion process in non-Euclidean spaces, especially when modeling hierarchical structures in any graphs and manifolds.

\subsubsection{Poincaré Normal Distribution's Non-Additivity.} 
Here, we follow the approach of the paper~\cite{fu2024hyperbolic} to further discuss the non-additivity of the Poincaré normal distribution.

In anisotropic environments or settings, where properties vary depending on direction, the probability density of the phenomenon in question can be mathematically expressed using the following equation: 
\begin{equation}
    \begin{aligned}
\mathcal{N}_{\mathcal{B}_\kappa^d}^\text{P}(z|\mu, \Sigma) = \mathcal{N} \left(\lambda_\mu^\kappa \log_\mu(z) \Big| \mathbf{0}, \Sigma\right) \left(\frac{\sqrt{\kappa}d_p^\kappa({\boldsymbol{\mu}}, z)}{\sinh(\sqrt{\kappa}d_p^\kappa({\boldsymbol{\mu}}, z))}\right)^{d-1}.
\end{aligned}
\end{equation}

The density can be expressed by introducing the variable \( v = r \alpha = \lambda_{\mu}^{\kappa} \log_{\mu} (z) \) and utilizing the metric tensor, leading to the following expression:
\begin{equation}
\begin{aligned}
&\int_{\mathcal{B}_\kappa^d} \mathcal{N}_{\mathcal{B}_\kappa^d}^\mathbf{P}(z|\mu, \Sigma)d\mathcal{M}(z) \notag \\ 
=&\int_{\mathbb{R}^d} \mathcal{N}(v|\mathbf{0}, \Sigma) \left(\frac{\sqrt{\kappa}\|v\|_2}{\sinh(\sqrt{\kappa}\|v\|_2)}\right)^{d-1} \left(\frac{\sinh(\sqrt{\kappa}\|v\|_2)}{\sqrt{\kappa}\|v\|_2}\right)^{d-1} dv \notag \\ 
=&\int_{\mathbb{R}^d} \mathcal{N}(v|\mathbf{0}, \Sigma)\,dv.
\end{aligned}
\end{equation}

Next, the derivation is made to determine whether the sum of two independent Poincaré normally distributed variables still satisfies the Poincaré normal distribution:
\begin{equation}
\begin{aligned}
\mathcal{N}_{\mathcal{B}_\kappa^d}^\mathbf{P}&(z_1|\mu_1, \Sigma_1) * \mathcal{N}_{\mathcal{B}_\kappa^d}^\mathbf{P}(z_2|\mu_2, \Sigma_2) \\[6pt]
= \int_{\mathcal{B}_\kappa^d}& \mathcal{N}_{\mathcal{B}_\kappa^d}^\mathbf{P}(z - z_2|\mu_1, \Sigma_1)\mathcal{N}_{\mathcal{B}_\kappa^d}^\mathbf{P}(z_2|\mu_2, \Sigma_2)d\mathcal{M}(z_2) \\[6pt]
= \int_{\mathbb{R}^d}& \mathcal{N}(v - v_2|\mathbf{0}, \Sigma_1)\mathcal{N}(v_2|\mathbf{0}, \Sigma_2)\left(\frac{\sqrt{\kappa}\|v - v_2\|_2}{\sinh(\sqrt{\kappa}\|v - v_2\|_2)}\right)^{d-1} dv_2 \\[6pt] 
\end{aligned}
\end{equation}
Here, we know that the Poincaré normal distribution does not exhibit additivity in anisotropic environments:
\begin{equation}
\begin{aligned}
\mathcal{N}_{\mathcal{B}_\kappa^d}^\mathbf{P}&(z_1|\mu_1, \Sigma_1) * \mathcal{N}_{\mathcal{B}_\kappa^d}^\mathbf{P}(z_2|\mu_2, \Sigma_2) \not\sim \mathcal{N}_{\mathcal{B}_\kappa^d}^\mathbf{P}(z|\mu, \Sigma).
\label{35}
\end{aligned}
\end{equation}



On the other hand, in the isotropic setting, the density of the Poincaré normal distribution is given by:
\begin{equation}
\begin{split}
\mathcal{N}_{\mathcal{B}_\kappa}^\text{P}(z|\boldsymbol{\mu}, \sigma^2) = (2\pi\sigma^2)^{-d/2} \exp\left(-\frac{d_p^\kappa(\boldsymbol{\mu}, z)^2}{2\sigma^2}\right) \left(\frac{\sqrt{\kappa}d_p^\kappa(\boldsymbol{\mu}, z)}{\sinh(\sqrt{\kappa}d_p^\kappa(\boldsymbol{\mu}, z))}\right)^{d-1}.
\end{split}
\end{equation}

The integral form of this density is:
\begin{equation}
\int_{\mathcal{B}_\kappa^d} \mathcal{N}_{\mathcal{B}_\kappa}^\text{P}(z|\boldsymbol{\mu}, \sigma^2) d\mathcal{M}(z) = \int_{R_+} \int_{S^{d-1}} \frac{1}{Z^R} e^{-\frac{r^2}{2\sigma^2}} r^{d-1} \, dr \, ds_{S^{d-1}},
\end{equation}
where \( Z^R \) is the normalization constant, defined as:
\begin{equation}
Z^R = \lambda \binom{d-1}{k} e^{\frac{(d-1-2k)^2}{2}c\sigma^2} \left[1 + \text{erf}\left(\frac{(d-1-2k)\sqrt{c\sigma}}{\sqrt{2}}\right)\right],
\end{equation}
with the $\lambda$ is defined as:
\begin{equation}
\lambda = \frac{2\pi^{d/2}}{\Gamma(d/2)} \sqrt{\frac{\pi}{2}}\sigma \frac{1}{(2\sqrt{c})^{d-1}} \sum_{k=0}^{d-1} (-1)^k.
\end{equation}

Next, the additivity can be derived as follows:
\begin{equation}
\begin{aligned}
\mathcal{N}_{\mathcal{B}_\kappa^d}^\mathbf{P}&(z_1|\mu_1, \Sigma_1) * \mathcal{N}_{\mathcal{B}_\kappa^d}^\mathbf{P}(z_2|\mu_2, \Sigma_2) \\ 
= \int_{\mathcal{B}_\kappa^d}&\mathcal{N}_{\mathcal{B}_\kappa^d}^\mathbf{P}(z - z_2|\mu_1, \Sigma_1)\mathcal{N}_{\mathcal{B}_\kappa^d}^\mathbf{P}(z_2|\mu_2, \Sigma_2)d\mathcal{M}(z_2) \\ 
= \int_{R_+}& \int_{S^{d-1}} \frac{1}{Z^{R^2}}e^{-\frac{(r-r_2)^2}{2\sigma^2}}(r - r_2)^{d-1}\gamma_p^\kappa e^{-\frac{(r_2)^2}{2\sigma^2}}(r_2)^{d-1}drds_{S^{d-1}} \\ 
\end{aligned}
\end{equation}
\begin{equation}
\begin{aligned}
\Rightarrow\mathcal{N}_{\mathcal{B}_\kappa^d}^\mathbf{P}&(z_1|\mu_1, \Sigma_1) * \mathcal{N}_{\mathcal{B}_\kappa^d}^\mathbf{P}(z_2|\mu_2, \Sigma_2) \not\sim \mathcal{N}_{\mathcal{B}_\kappa^d}^\mathbf{P}(z|\mu, \Sigma).
\end{aligned}
\end{equation}

As stated in the conclusion of E.q. \eqref{35}, the Poincaré normal distribution does not exhibit additivity even in the isotropic setting.

\subsection{Theoretical Analysis of Hyperbolic Diffusion Distance}

Here, we provide a concise analysis of the hyperbolic diffusion distance (HDD)~\cite{lin2023hyperbolic}, focusing on its theoretical foundation and the relationship with hierarchical structures.
Briefly speaking, it approximates the geodesic distance on a Riemannian manifold with non-negative curvature, providing a natural metric for hierarchical data structures.

\subsubsection{HDD Recovers Hierarchical Distance Without Explicit Tree Structure}
To obtain this result, we need to prove that \( d_{\text{HDD}} \) and \( d_{T}^{2\alpha} \) are equivalent under certain conditions, as expressed by the following formula:
\begin{equation}
d_{\text{HDD}} \Leftrightarrow d_{T}^{2\alpha}: \quad \text{for} \; 0 < \alpha \leq \frac{1}{2}, \, \text{ and } K, n \text{ are sufficiently large}.
\end{equation}
This equation shows that HDD recovers the hierarchical distance even without explicit tree structure information. Practically, \( \alpha \) should be set close to \( \frac{1}{2} \) for better approximation of the hierarchical distance. As \( \alpha \to \frac{1}{2} \), \( d_{\text{HDD}} \) approximates the 0-hyperbolic distance\cite{kharlampovich1998hyperbolic}. The detailed proof details can be refered to paper~\cite{lin2023hyperbolic}.

\subsubsection{Geometric Measures in Graphs}
In the context of graph-based analysis, geometric measures play a crucial role in understanding the relationships between nodes. Two important types of measures are the shortest path metric and the multi-scale metric in continuous space, which are fundamental for various graph-related tasks such as diffusion models, hierarchical clustering, and graph-based learning.

\subsubsection{Shortest Path Metric and Its Significance}
The shortest path metric \(d_T(u, v)\) represents the length of the shortest path between two nodes \(u\) and \(v\). This metric is of great importance, especially in diffusion models and hierarchical clustering in graphs. In tree-like structures, it captures the most efficient way to traverse between nodes. By leveraging the shortest path metric, we can better understand the topological structure of the graph and how information spreads within it.

\subsubsection{Multi-Scale Metric in Continuous Space}
\paragraph{Local Geometric Measure Definition}
The local geometric measure at scale \(k\) is defined using the unnormalized Hellinger distance between probability distributions. The formula is given as:
\begin{equation}
\begin{split}
M_k(x, x')&=\sqrt{\left(\sqrt{a_{2 - k}(x, \cdot)} - \sqrt{a_{2 - k}(x', \cdot)}\right)^T\left(\sqrt{a_{2 - k}(x, \cdot)} - \sqrt{a_{2 - k}(x', \cdot)}\right)}\\&=\sqrt{\sum_{i}\left(\sqrt{a_{2 - k}(x, i)} - \sqrt{a_{2 - k}(x', i)}\right)^2}.
\end{split}
\end{equation}

This measure provides a local view of the geometric relationship between points \(x\) and \(x'\) at a specific scale \(k\).

\paragraph{Multi-Scale Metric Definition}
Based on the local geometric measure, the multi-scale metric is defined using the inverse hyperbolic sine function of the scaled Hellinger measure. The general form of the multi-scale metric is:
\begin{equation}
M_k(x, x') = \left\| \sqrt{a_{2 - k}(x, \cdot)} - \sqrt{a_{2 - k}(x', \cdot)} \right\|_2,
\end{equation}
where \(0 < \alpha < 1\). This metric combines the local geometric information at different scales \(k\) to provide a more comprehensive view of the geometric relationship between points.

\paragraph{Approximation of the Multi-Scale Metric}
In practice, the multi-scale metric \(\hat{M}_\alpha(x, x')\) can be approximated by the first \(K\) terms. The approximation formula is:
\begin{equation}
\hat{M}_\alpha(x, x')\approx \sum_{k = 0}^{K} 2\sinh^{-1} \left( e^{(1 - k\alpha)\ln(2)}M_k(x, x') \right).
\end{equation}

This approximation simplifies the calculation of the multi-scale metric while still retaining a significant amount of information.

\subsubsection{Conclusion: The Role of HDD in Hierarchical Structure Recovery}
The Hyperbolic Diffusion Distance provides an effective way to recover hierarchical structures from data using hyperbolic geometry. By optimizing hyperbolic embeddings with the multi-scale metric, HDD can capture the hierarchical relationships between data points. This property makes HDD applicable in various tasks such as clustering and graph-based learning, where understanding the hierarchical structure of the data is essential.

\begin{figure*}[h]
	 \centering
	\begin{minipage}{0.32\linewidth}
		\vspace{10pt}
\centerline{\includegraphics[width=1.0\textwidth]{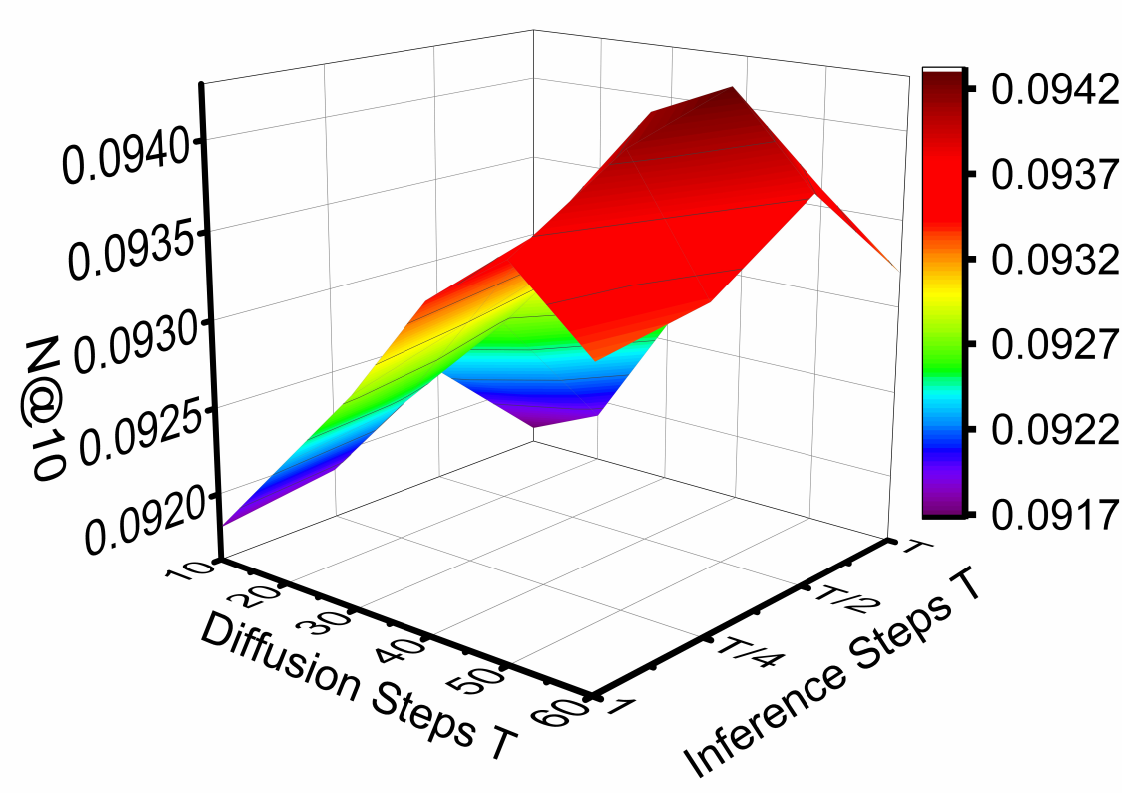}}
		\centerline{(a) ML-1M}
	\end{minipage}
	\begin{minipage}{0.32\linewidth}
		\vspace{10pt}
\centerline{\includegraphics[width=1.0\textwidth]{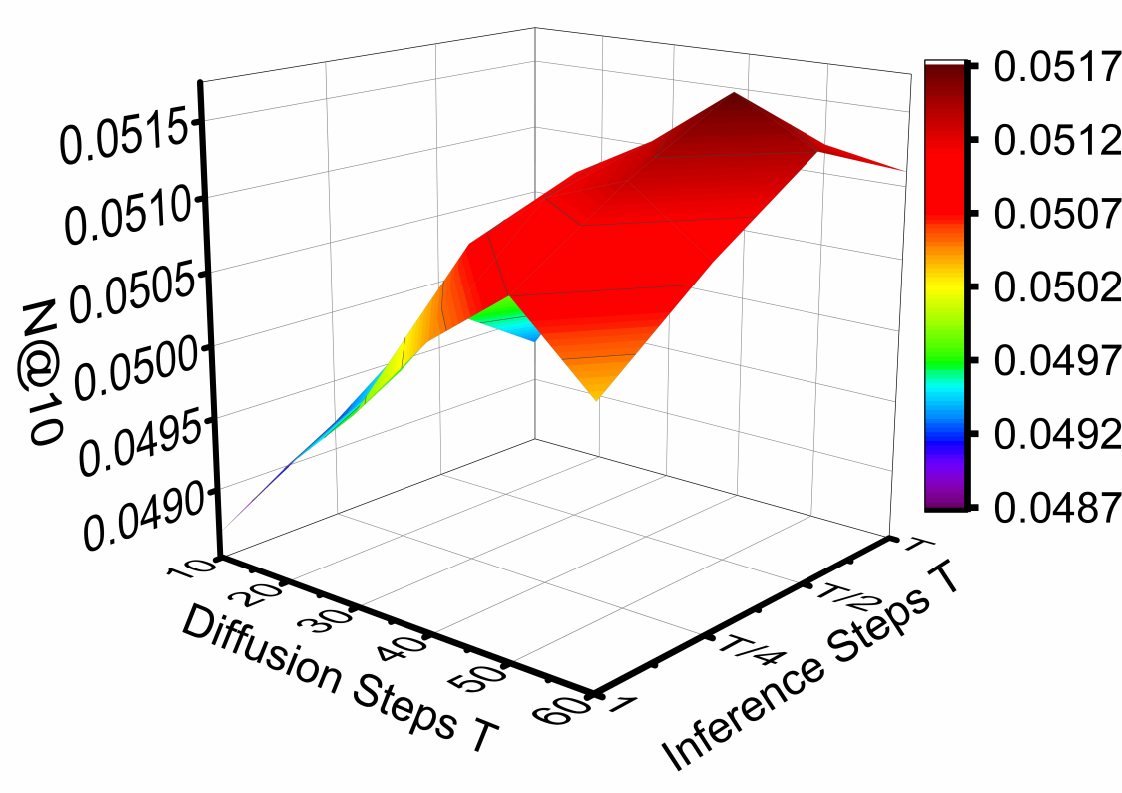}}
		\centerline{(b) Amazon-Book}
	\end{minipage}
 	\begin{minipage}{0.32\linewidth}
		\vspace{10pt}
\centerline{\includegraphics[width=1.0\textwidth]{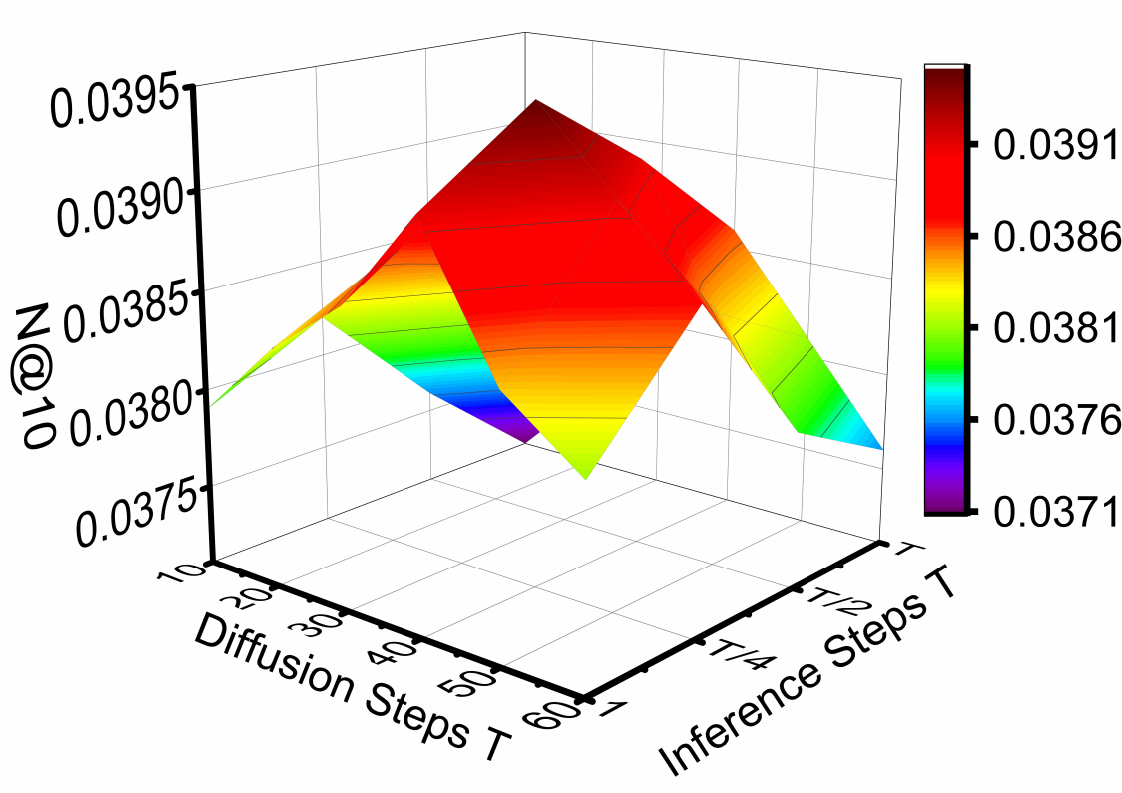}}
		\centerline{(c) Yelp2020}
	\end{minipage}
   \caption{The variation of model performance across three datasets as diffusion steps and inference steps change.}
	\label{fig_step_n10}
\end{figure*}

\section{Experiments}

\subsection{Experimental Settings}
\subsubsection{Baselines}
\label{baselines}
The detailed information of the baselines is as follows:

\noindent \underline{\textbf{Classic Collaborative Filtering Methods:}}
\begin{itemize}[leftmargin=*]
    \item \textbf{BPRMF}~\cite{rendle2009bpr}: This is a typical collaborative filtering method that optimizes MF with a pairwise ranking loss.
    \item \textbf{LightGCN}~\cite{he2020lightgcn}: This is an effective GCN-based collaborative filtering method, which improves performance by eliminating non-linear projection and activation.
\end{itemize}

\noindent \underline{\textbf{Auto-Encoders Recommender Methods:}}
\begin{itemize}[leftmargin=*]
    \item \textbf{CDAE}~\cite{wu2016collaborative}: This is a collaborative filtering method that applies denoising auto-encoders with user-specific latent factors to improve top-N recommendation performance.

    \item \textbf{MultiDAE}~\cite{liang2018variational}: This is a variational autoencoder approach with partial regularization and multinomial likelihood for collaborative filtering on implicit feedback data

\end{itemize}

\noindent \underline{\textbf{Diffusion Recommender Methods:}}
\begin{itemize}[leftmargin=*]
    \item \textbf{CODIGEM}~\cite{walker2022recommendation}: This method employs a simple CL approach that avoids graph augmentations and introduces uniform noise into the embedding space to generate contrastive views.
    \item \textbf{DiffRec}~\cite{wang2023diffusion}: This method uses LightGCN as the backbone and incorporates a series of structural augmentations to enhance representation learning.
    \item 
    \textbf{DDRM}~\cite{zhao2024denoising}: This is a plug-in denoising diffusion model that enhances robust representation learning for existing recommender systems by iteratively injecting and removing noise from user and item embeddings.

\end{itemize}
\noindent \underline{\textbf{Hyperbolic Recommender Methods:}}
\begin{itemize}[leftmargin=*]
    \item \textbf{HyperML}~\cite{vinh2020hyperml}: This method is the first to propose using hyperbolic margin ranking loss for predicting user preferences toward items.
    \item \textbf{HGCF}~\cite{sun2021hgcf}: This method is the first hyperbolic GCN model for collaborative filtering that can be effectively learned using a margin ranking loss.
    \item \textbf{HICF}~\cite{yang2022hicf}: This method adapts hyperbolic margin ranking learning by making the pull and push procedures geometric-aware, aiming to provide informative guidance for the learning of both head and tail items.
\end{itemize}

\begin{figure*}
	 \centering
	\begin{minipage}{0.32\linewidth}
		\vspace{10pt}
\centerline{\includegraphics[width=1.0\textwidth]{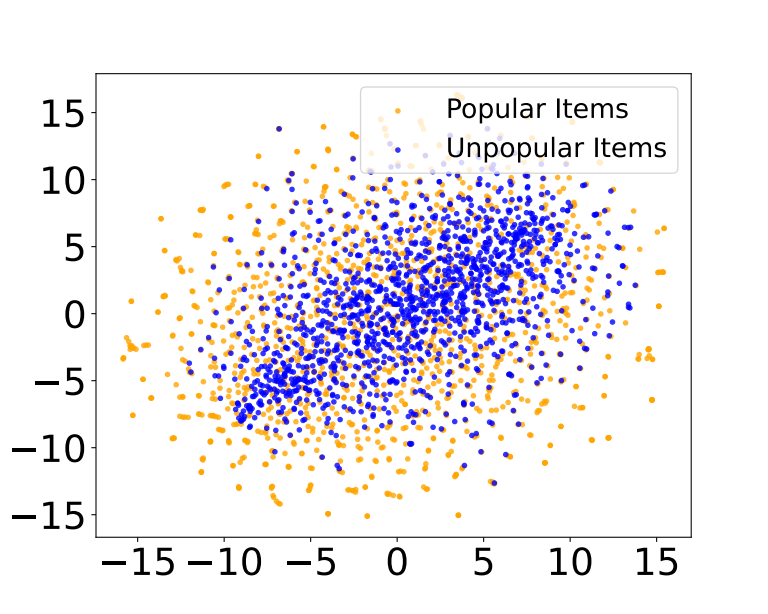}}
		\centerline{DDRM}
	\end{minipage}
 \begin{minipage}{0.32\linewidth}
		\vspace{10pt}
\centerline{\includegraphics[width=1.0\textwidth]{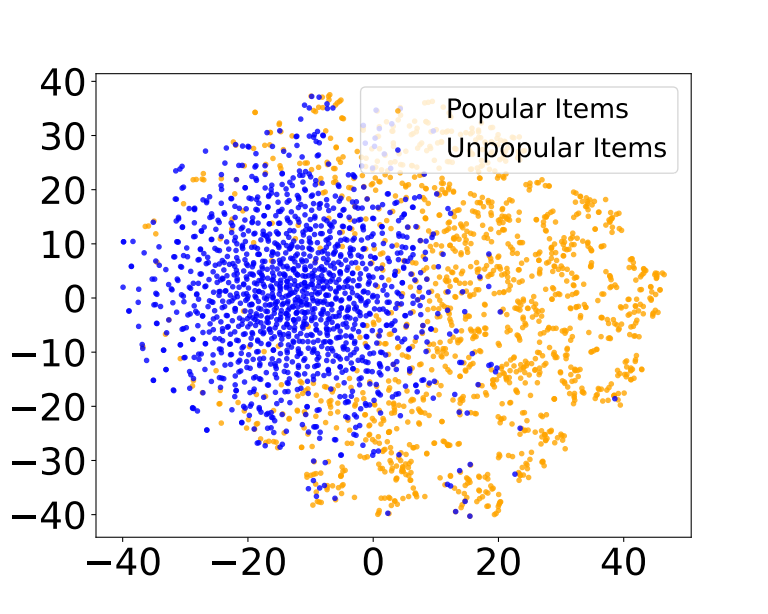}}
		\centerline{HICF}
	\end{minipage}
 \begin{minipage}{0.32\linewidth}
        \vspace{10pt}
\centerline{\includegraphics[width=1.\textwidth]{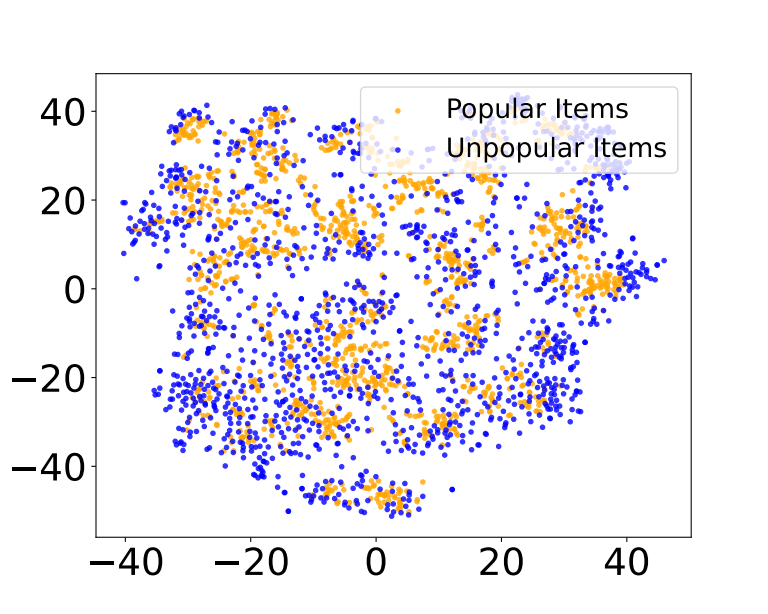}}
        \centerline{HDRM}
    \end{minipage}
   \caption{Visualize the distribution of item embeddings on the ML-1M dataset using DDRM, HICF, and HDRM. HDRM ensures that popular and unpopular items have representations with almost the same positions in the same space.
.}
	\label{fig_emb}
\end{figure*}
\begin{figure*}
	 \centering
	\begin{minipage}{0.32\linewidth}
		\vspace{10pt}
\centerline{\includegraphics[width=1.0\textwidth]{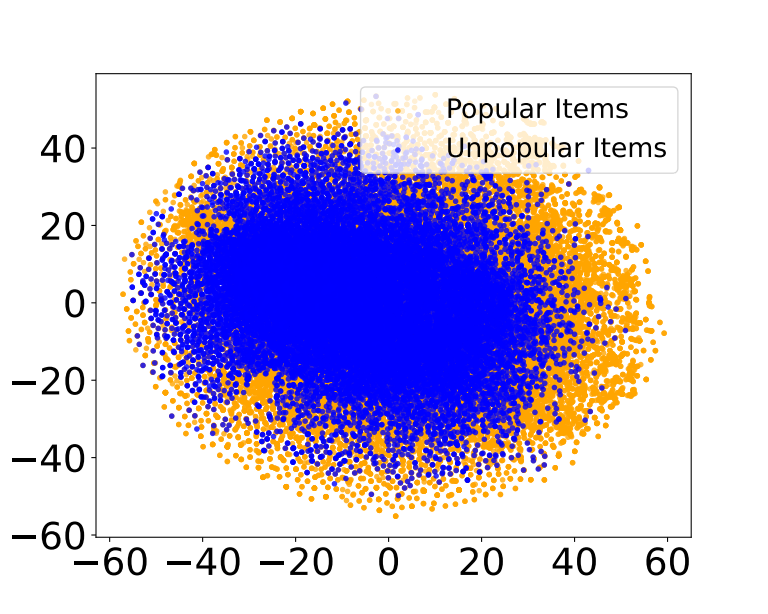}}
		\centerline{DDRM}
	\end{minipage}
 \begin{minipage}{0.32\linewidth}
		\vspace{10pt}
\centerline{\includegraphics[width=1.0\textwidth]{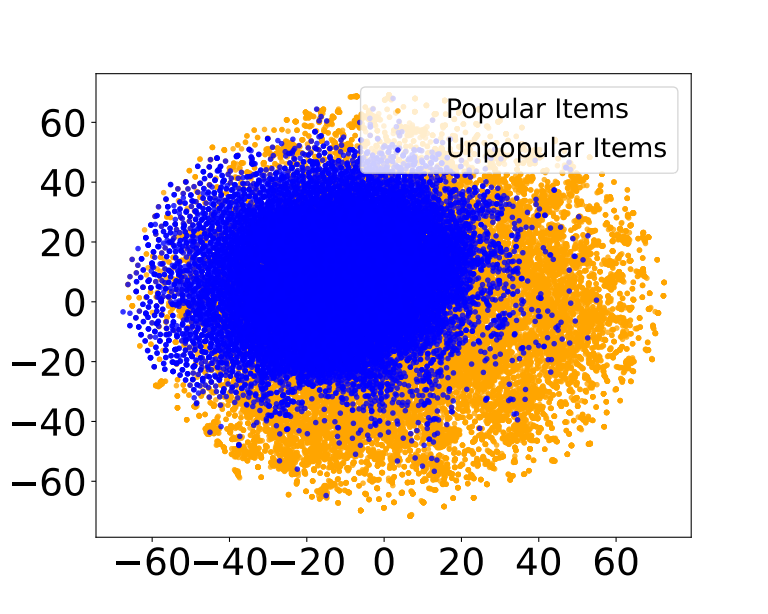}}
		\centerline{HICF}
	\end{minipage}
 \begin{minipage}{0.32\linewidth}
        \vspace{10pt}
\centerline{\includegraphics[width=1.\textwidth]{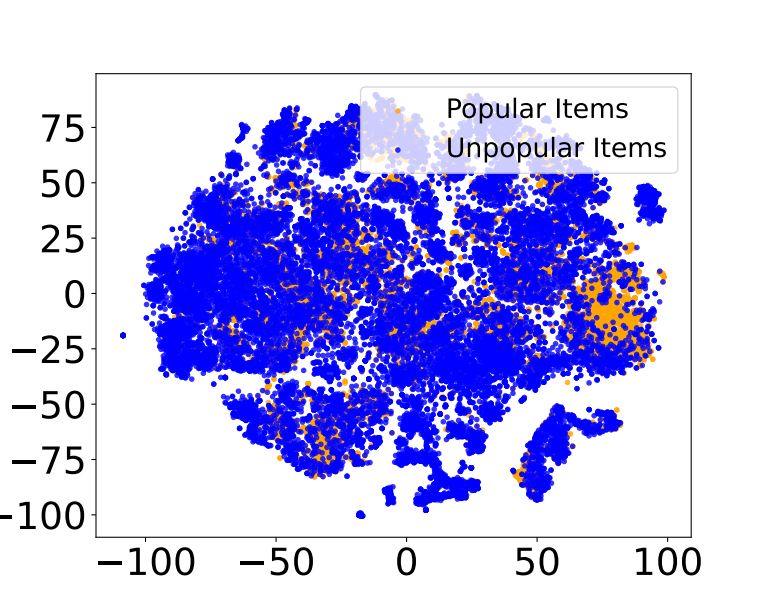}}
        \centerline{HDRM}
    \end{minipage}
   \caption{Visualize the distribution of item embeddings on the Amazon-Book dataset using DDRM, HICF, and HDRM. HDRM ensures that popular and unpopular items have representations with almost the same positions in the same space.
}
	\label{fig_emb1}
\end{figure*}
\begin{figure*}
	 \centering
	\begin{minipage}{0.32\linewidth}
		\vspace{10pt}
\centerline{\includegraphics[width=1.0\textwidth]{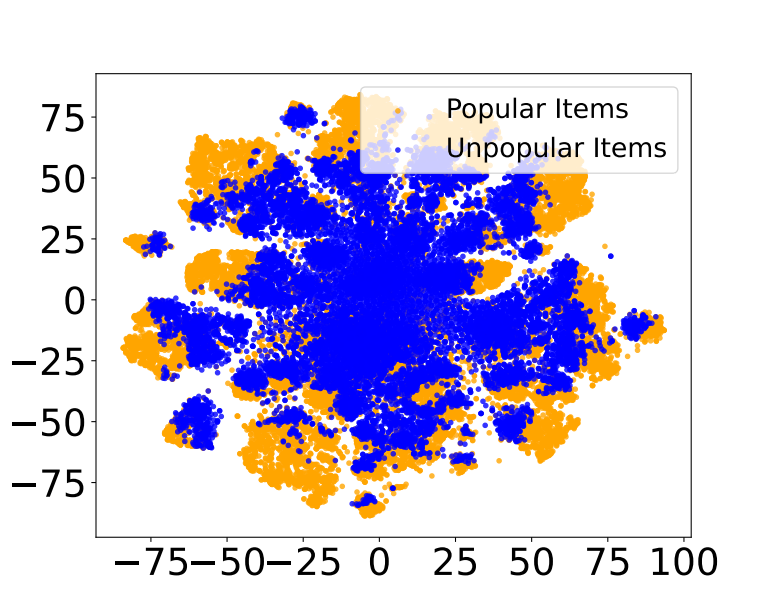}}
		\centerline{DDRM}
	\end{minipage}
 \begin{minipage}{0.32\linewidth}
		\vspace{10pt}
\centerline{\includegraphics[width=1.0\textwidth]{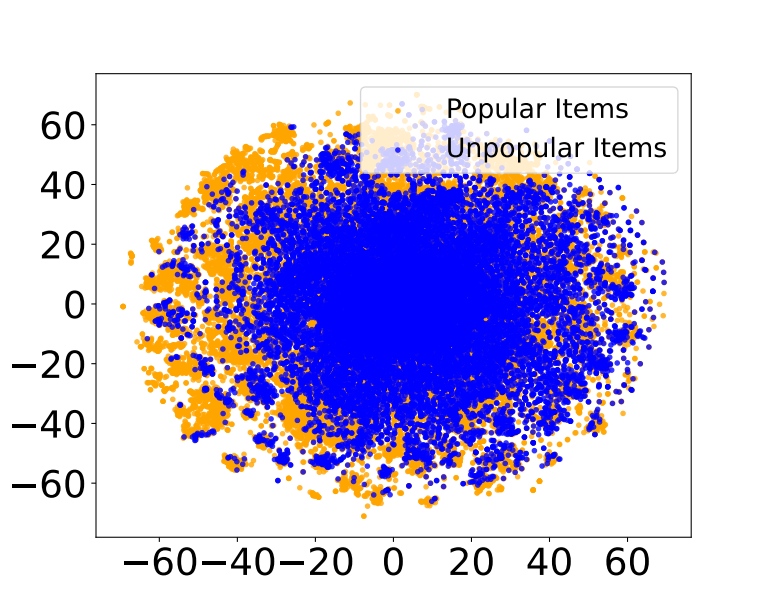}}
		\centerline{HICF}
	\end{minipage}
 \begin{minipage}{0.32\linewidth}
        \vspace{10pt}
\centerline{\includegraphics[width=1.\textwidth]{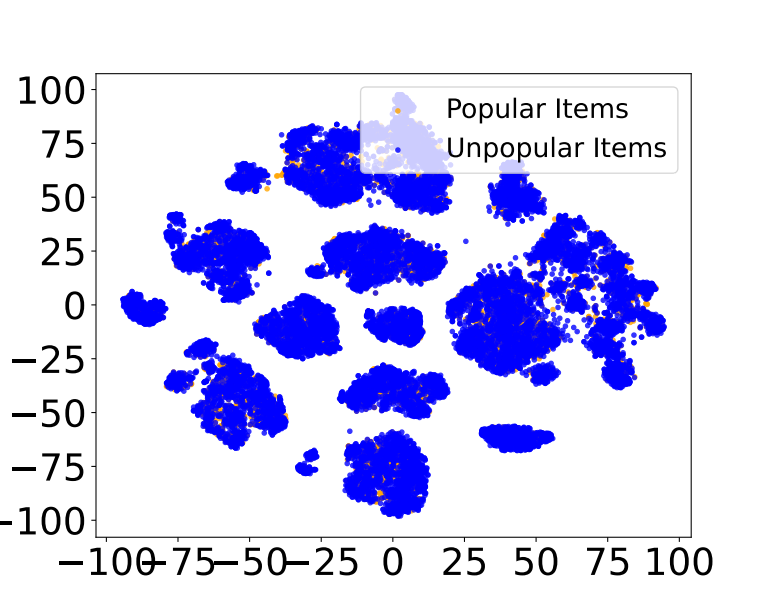}}
        \centerline{HDRM}
    \end{minipage}
   \caption{Visualize the distribution of item embeddings on the Yelp2020 dataset using DDRM, HICF, and HDRM. HDRM ensures that popular and unpopular items have representations with almost the same positions in the same space.
}
	\label{fig_emb2}
\end{figure*}
\subsubsection{Hyper-parameter Settings}
\label{hyperparameter}
We determine the optimal hyperparameters based on the Recall@20 metric evaluated on the validation set. For our Hyperbolic model, we tune these key parameters. The learning rate is varied among $\{ 1 e^{-4},5 e^{-4},1 e^{-3},5 e^{-3} \}$
, while the curvature $\kappa$ is set to either -1 or 1. We explore GCN architectures with $\{2, 3, 4\}$ layers, and weight decay values of $\{0.001, 0.005, 0.01\}$. The margin is tested at $\{0.05, 0.1, 0.15, 0.2, 0.25, 0.3, 0.35, 0.4\}$. For the diffusion model, we investigate diffusion steps $T$ ranging from $\{10, 20, 30, 40, 50, 60\}$. The noise schedule is bounded between $1 e^{-4}$ and $1 e^{-2}$. We explore loss balance factors $\alpha$ 
 from $\{0.1, 0.2, ..., 0.6\}$, and reweighted factors $\gamma$ from $\{0, 0.05, 0.1, 0.2, ... , 0.9\}$. All experiments are conducted using PyTorch on a server equipped with 16 Intel Xeon CPUs @2.10GHz and an NVIDIA RTX 4090 GPU, ensuring efficient training and evaluation of our models across this extensive hyperparameter space.
 
\subsection{More Experimental Results}
\subsubsection{Diffusion Step Analysis}
\label{n10}
Here is further analysis of the diffusion steps. Figure \ref{fig_step_n10} illustrates how the model performance metric N@10 changes across three datasets as the number of diffusion and inference steps vary. During the diffusion process, the model gradually spreads information across different nodes or features in the data. With more diffusion steps, the model can capture more complex relationships and patterns in the data. In recommender systems, it can better understand the relationships between users and items, and thus make more accurate recommendations. The inference steps, on the other hand, help the model refine these relationships and generate more reliable predictions.

However, this improvement in performance does not continue indefinitely. There exists a certain threshold beyond which the performance of the HDRM sharply declines. This phenomenon can be attributed to several factors. One possible reason is over-exploration. As the number of diffusion and inference steps increases, the model may begin to explore irrelevant or noisy parts of the data space. This can lead to the model being overly influenced by outliers or random fluctuations in the data, resulting in less accurate predictions. Another factor could be the computational complexity. With a large number of diffusion and inference steps, the computational cost of the model increases significantly. This may lead to longer training and inference times, and in some cases, memory issues. As a result, the model's performance may degrade due to resource limitations.

In conclusion, when optimizing the HDRM, it is crucial to find the optimal number of diffusion and inference steps. This requires a careful balance between exploring the data space to capture complex relationships and avoiding over-exploration and excessive computational costs. Future research could focus on developing more sophisticated methods to automatically determine the optimal number of steps based on the characteristics of the dataset.

\subsubsection{Embedding Visualization}
Figures \ref{fig_emb}, \ref{fig_emb1}, and \ref{fig_emb2} present t-SNE visualizations of item embeddings learned by DDRM, HICF, and HDRM on the ML-1M, Amazon-Book, and Yelp2020 datasets, offering insights into our model's capability to address distribution shifts. We categorize items based on their popularity in the training set. For ML-1M and Yelp2020, the top 50\% most popular items are designated as "popular", while the bottom 50\% are labeled "unpopular". Due to its larger size, the Amazon-Book dataset uses a 20-80 split for popular and unpopular items, respectively.

The visualizations reveal that DDRM's learned embeddings for popular and unpopular items maintain a noticeable separation in the representation space. In contrast, HDRM achieves a more uniform distribution of both types of embeddings within the same space. This observation suggests that HDRM effectively mitigates the tendency of recommender systems to over-recommend popular items at the expense of niche selections.
Interestingly, HICF demonstrates a more pronounced differentiation between the two embedding categories. This characteristic can be attributed to the curvature of hyperbolic space, which allows for exponential growth of representational capacity within a finite area. Consequently, this property naturally amplifies item distinctions, particularly in terms of popularity.

\end{document}